\documentclass[12pt, notitlepage]{article}

\usepackage{a4}

\usepackage{xcolor}
\usepackage{algorithm}
\usepackage[noend]{algpseudocode}

\definecolor{TUMblue}{RGB}{0, 101, 189}
\definecolor{TUMlightblue}{RGB}{100,160,200}
\definecolor{TUMgreen}{RGB}{162,173,0}
\definecolor{TUMorange}{RGB}{227,114,034}
\definecolor{TUMivory}{RGB}{218,215,203}

\usepackage{natbib}

\usepackage{hyperref}
\hypersetup{
colorlinks=true,
linkcolor=TUMblue,
citecolor=TUMblue,
filecolor=TUMblue,
urlcolor=TUMblue
}

\usepackage{etoolbox}
\makeatletter

\pretocmd{\NAT@citex}{%
\let\NAT@hyper@\NAT@hyper@citex
\def\NAT@postnote{#2}%
\setcounter{NAT@total@cites}{0}%
\setcounter{NAT@count@cites}{0}%
\forcsvlist{\stepcounter{NAT@total@cites}\@gobble}{#3}}{}{}
\newcounter{NAT@total@cites}
\newcounter{NAT@count@cites}
\def\NAT@postnote{}

\def\NAT@hyper@citex#1{%
\stepcounter{NAT@count@cites}%
\hyper@natlinkstart{\@citeb\@extra@b@citeb}#1%
\ifnumequal{\value{NAT@count@cites}}{\value{NAT@total@cites}}
{\ifNAT@swa\else\if*\NAT@postnote*\else%
\NAT@cmt\NAT@postnote\global\def\NAT@postnote{}\fi\fi}{}%
\ifNAT@swa\else\if\relax\NAT@date\relax
\else\NAT@@close\global\let\NAT@nm\@empty\fi\fi% avoid compact citations
\hyper@natlinkend}
\renewcommand\hyper@natlinkbreak[2]{#1}

\patchcmd{\NAT@citex}
{\ifNAT@swa\else\if*#2*\else\NAT@cmt#2\fi
\if\relax\NAT@date\relax\else\NAT@@close\fi\fi}{}{}{}
\patchcmd{\NAT@citex}
{\if\relax\NAT@date\relax\NAT@def@citea\else\NAT@def@citea@close\fi}
{\if\relax\NAT@date\relax\NAT@def@citea\else\NAT@def@citea@space\fi}{}{}

\makeatother

\usepackage{amssymb, graphicx}
\usepackage{subfigure}
\usepackage{amsmath}
\usepackage{amsthm}
\usepackage{undertilde}
\usepackage{verbatim}
\usepackage{bbm}
\usepackage{ dsfont }
\usepackage{geometry}
\usepackage{pdflscape}
\usepackage{multirow}
\usepackage{aliascnt}

\newcommand{\mynewtheorem}[2]{
\newaliascnt{#1}{dummy}
\newtheorem{#1}[#1]{#2}
\aliascntresetthe{#1}
\expandafter\def\csname #1autorefname\endcsname{#2}
}

\theoremstyle{definition}
\mynewtheorem{thm}{Theorem}
\mynewtheorem{defi}{Definition}%[section]
\mynewtheorem{lem}{Lemma}%[section]
\mynewtheorem{cor}{Corollary}%[section]
\mynewtheorem{prop}{Proposition}%[section]
\mynewtheorem{exa}{Example}%[section]
\mynewtheorem{alg}{Algorithm}%[section]
\mynewtheorem{rem}{Remark}%[section]
\mynewtheorem{bsp}{Example}

\def\equationautorefname~#1\null{Equation~(#1)\null}

\newcommand{\aref}[1]{\hyperref[#1]{Appendix~\ref{#1}}}

\usepackage{sectsty}
\allsectionsfont{\sffamily}

\usepackage{url}

\usepackage{caption}
\usepackage{footnote}

\newcommand{\Ebb}{\mathbb{E}}

\newcommand{\rb}{\mathbf{r}}

\newcommand{\ub}{\mathbf{u}}
\newcommand{\Ub}{\mathbf{U}}

\newcommand{\vb}{\mathbf{v}}

\newcommand{\wb}{\mathbf{w}}

\newcommand{\xb}{\mathbf{x}}

\newcommand{\Dc}{\mathcal{D}}

\newcommand{\Ic}{\mathcal{I}}

\newcommand{\Rc}{\mathcal{R}}

\newcommand{\be}{\begin{equation}}
\newcommand{\ee}{\end{equation}}

\newcommand{\eps}{\varepsilon}

\DeclareMathOperator{\argmax}{arg\,max}

\DeclareMathOperator{\dKL}{dKL}

\DeclareMathOperator{\KL}{KL}

\DeclareMathOperator{\sdKL}{sdKL}

\newcommand*\diff{\mathop{}\!\mathrm{d}}

\begin{document}
	
	{
		\renewcommand*{\thefootnote}{\fnsymbol{footnote}}
		\title{\textbf{\sffamily Using model distances to investigate\\the simplifying assumption, model selection \\and truncation levels for vine copulas			 
			}}
			
			\date{\small \today}
			\author{Matthias Killiches$^*$ \and Daniel Kraus\footnote{Zentrum Mathematik, Technische Universit\"at M\"unchen, Boltzmannstra\ss e 3, 85748 Garching, Germany} \footnote{Corresponding author, email: \texttt{daniel.kraus@tum.de}.}  \and Claudia Czado$^*$}
			
			\maketitle
			\vspace*{-5mm}
			\begin{abstract}
				Vine copulas are a useful statistical tool to describe the dependence structure between several random variables, especially when the number of variables is very large. When modeling data with vine copulas, one often is confronted with a set of candidate models out of which the best one is supposed to be selected. For example, this may arise in the context of non-simplified vine copulas, truncations of vines and other simplifications regarding pair-copula families or the vine structure. With the help of distance measures we develop a parametric bootstrap based testing procedure to decide between copulas from nested model classes. In addition we use distance measures to select among different candidate models. All commonly used distance measures, e.g.\ the Kullback-Leibler distance, suffer from the curse of dimensionality due to high-dimensional integrals. As a remedy for this problem, Killiches, Kraus and Czado (2017b) propose several modifications of the Kullback-Leibler distance. We apply these distance measures to the above mentioned model selection problems and substantiate their usefulness.
				\\
				
			\end{abstract}
			%\vspace{5mm}
		}
%\tableofcontents

\section{Introduction}\label{sec:intro}
In a world of growing data sets and rising computational power the need of adequately modeling multivariate random quantities is self-evident. Since the seminal paper of \cite{Sklar} the modeling of a multivariate distribution function can be divided into separately considering the marginal distributions and the underlying dependence structure, the so-called \emph{copula}. One of the most popular copula classes, especially for high-dimensional data, are \emph{vine copulas} \citep{aasczado}. Constructing a multivariate copula in terms of bivariate building blocks, this pair-copula construction has the advantage of being highly flexible while still yielding interpretable models. Vines have been extensively used for high-dimensional copula modeling. \cite{brechmann2013risk} analyzed the interdependencies of the stocks contained in the Euro Stoxx 50 for risk management purposes, while \cite{stoeber2012detecting} considered the detection of regime switches in high-dimensional financial data. \cite{muller2016representing} used Gaussian directed acyclic graphs to facilitate the estimation of vine copulas in very high dimensions. Moreover, in the context of big data, gene expression data and growing market portfolios, the interest in high-dimensional data modeling cannot be expected to decline.

In model selection, the distance between statistical models plays a big role. Usually the difference between two statistical models is measured in terms of the Kullback-Leibler (KL) distance \citep{kullback1951information}. Model selection procedures based on the KL distance for copulas are developed for example in \cite{chen2005pseudo}, \cite{chen2006estimation} and \cite{diks2010out}. In the context of vine copulas, \cite{joe2014dependence} used the KL distance to calculate the sample size necessary to discriminate between two densities. Investigating the simplifying assumption \cite{haff2010simplified} used the KL distance to find the simplified vine closest to a given non-simplified vine and \cite{stoeber2013simplified} assessed the strength of non-simplifiedness of the trivariate Farlie-Gumbel-Morgenstern (FGM) copula for different dependence parameters.

Nevertheless, the main issue of the Kullback-Leibler distance is that, as soon as it cannot be computed analytically, a numerical evaluation of the appearing integral is needed, which is hardly tractable once the model dimension exceeds three or four. In order to tackle this problem, several modifications of the Kullback-Leibler distance have been proposed in \cite{killiches2015model}. They yield model distances which are close in performance to the classical KL distance, however with much faster computation times, facilitating their use in very high dimensions. While the examples presented in that paper are mainly plausibility checks underlining the viability of the proposed distance measures, the aim of this paper is to demonstrate their usefulness in three practical problems. First, we investigate a major question arising when working with vines: Is the simplifying assumption justified for a given data set or do we need to account for non-simplifiedness? The importance of this topic can be seen from many recent publications such as \cite{haff2010simplified}, \cite{stoeber2012}, \cite{acar2012beyond}, \cite{spanhel2015simplified} or \cite{killiches2016examination}. Then we show how to select the best model out of a list of candidate models with the help of a model distance based measure. Finally, we also use the new distance measures to answer the question how to determine the optimal truncation level of a fitted vine copula, a task already recently discussed by \cite{brechmann2012truncated} and \cite{brechmann2015truncation}. Truncation methods have the aim of enabling high-dimensional vine copula modeling by severely reducing the number of used parameters without changing the fit of the resulting model too much.  

The remainder of the paper is structured as follows. In \autoref{sec:theory} we shortly introduce vine copulas and the distance measures proposed in \cite{killiches2015model}. Further, we provide a hypothesis test facilitating model selection. In \autoref{sec:simplified} we show how this test can be used to decide between simplified and non-simplified vines. \autoref{sec:model_selection} describes how the proposed distance measures can be applied to assess the best model fit out of a set of candidate models. As a final application the determination of the optimal truncation level of a vine copula is discussed in \autoref{sec:truncation}. Finally, \autoref{sec:conclusion} concludes.

\section{Theoretical concepts}\label{sec:theory}
\subsection{Vine copulas}\label{sec:vinecopulas}

\emph{Copulas} are $d$-dimensional distribution functions on $[0,1]^d$ with uniformly distributed margins. The usefulness of the concept of copulas has become clear with the publication of \cite{Sklar}, where the famous \emph{Sklar's Theorem} is proven. It provides a link between an arbitrary joint distribution and its marginal distributions and dependence structure. This result has been very important for applications since it allows the marginals and the dependence structure to be modeled separately. An introduction to copulas can be found in \cite{nelsen2006introduction}; \cite{joe1997multivariate} also contains a thorough overview over copulas.
%For any joint distribution function $F\colon \Rbb^d \to [0,1]$ of a $d$-dimensional random variable $(X_1,\ldots,X_d)'$ with univariate marginal distribution functions $F_j$, $j=1,\ldots,d$, there exists a copula $C$ such that
%\begin{equation}\label{eq:sklar}
%F(x_1,\ldots,x_d)=C\left(F_1(x_1),\ldots, F_d(x_d)\right).
%\end{equation}
%Further, if all $X_j$ are continuous random variables, this copula $C$ is unique. Assuming the existence of the so-called \emph{copula density} 
%\[
%c(u_1,\ldots,u_d):=\frac{\partial^d}{\partial u_1 \cdots \partial u_d}C(u_1,\ldots,u_d),
%\]
%one has 
%\[
%f(x_1,\ldots,x_d)=c\left(F_1(x_1),\ldots, F_d(x_d)\right)f_1(x_1)\cdots f_d(x_d),
%\]
%where $f_j$ are the marginal densities. In the following we will always assume absolute continuity of $C$ and the existence of $c$. \autoref{eq:sklar} can also be used to define a multivariate distribution by combining a copula $C$ and marginal distribution functions $F_j$. 
%Thus, the marginals and the dependence structure can be modeled separately, as we can specify the copula $C$ independently of the marginal distributions. 

Although there is a multitude of multivariate copula families (e.g.\ Gaussian, t, Gumbel, Clayton and Joe copulas), these models exhibit little flexibility in higher dimensions. Introducing \textit{vine copulas}, \cite{bedford2002vines} proposed a way of constructing copula densities by combining bivariate building blocks. \cite{aasczado} applied the concept of vines, also referred to as \textit{pair-copula constructions} (PCCs), and used them for statistical inference.

In the following we consider a $d$-dimensional random vector $\Ub=(U_1,\ldots,U_d)^\top$ with uniform marginals $U_j$, $j=1,\ldots,d$, following a copula $C$ with corresponding copula density $c$. For $j\in\left\{1,\ldots,d\right\}$ and $D\subseteq \left\{1,\ldots,d\right\}\setminus \left\{j\right\}$ we denote by $C_{j|D}$ the conditional distribution function of $U_j$ given $\Ub_D=(U_i)_{{i\in D}}$. For $j,k\in\left\{1,\ldots,d\right\}$ and $D\subseteq \left\{1,\ldots,d\right\}\setminus \left\{j,k\right\}$ the copula density of the distribution associated with the conditioned variables $U_j$ and $U_k$ given the conditioning variables $\Ub_D$ is denoted by $c_{j,k;D}$.

The structure of a $d$-dimensional vine copula is organized by a sequence of trees $\mathcal{V}=(T_1,\ldots,T_{d-1})$ satisfying
\begin{enumerate}
	\item $T_1=(V_1,E_1)$ is a tree with nodes $V_1=\left\{1,\ldots,d\right\}$ and edges $E_1$;
	\item For $m=2,\ldots,d-1$, the tree $T_m$ consists of nodes $V_m=E_{m-1}$ and edges $E_m$;
	\item Whenever two nodes of $T_m$ are connected by an edge, the corresponding edges of $T_{m-1}$ share a node ($m=2,\ldots,d-1$).
\end{enumerate}

In a vine copula model each edge of the $d-1$ trees corresponds to a bivariate pair-copula. Let $\bigcup_{m=1}^{d-1}\left\{ c_{j_e,k_e;D_e}\mid e\in E_m \right\}$ be the set of pair-copulas associated with the edges in $\mathcal{V}$, where -- following the notation of \cite{czado2010pair} -- $j_e$ and $k_e$ denote the indices of the conditioned variables $U_{j_e}$ and $U_{k_e}$ and $D_e$ represents the conditioning set corresponding to edge $e$. The vine density can be written as
\begin{equation}\label{eq:vine}
c(u_1,\ldots,u_d)=\prod_{m=1}^{d-1}\prod_{e\in E_m}c_{j_e,k_e;D_e}\left(C_{j_e|D_e}(u_{j_e}|\ub_{D_e}), C_{k_e|D_e}(u_{k_e}|\ub_{D_e});\ub_{D_e} \right). 
\end{equation}

\begin{figure}[h!]
	\centering
	\includegraphics[width=0.6\textwidth]{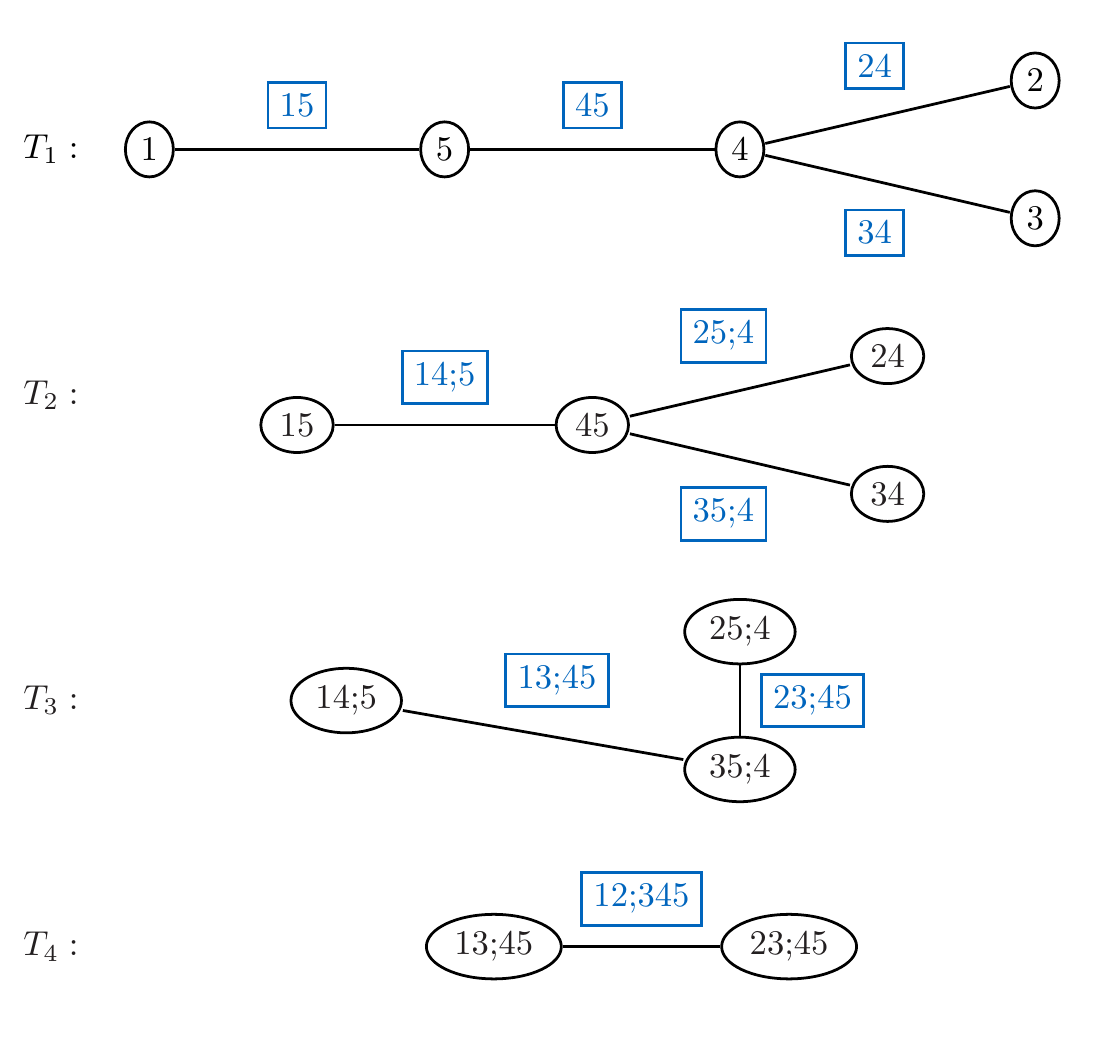}
	\caption{Tree sequence of an exemplary five-dimensional vine copula.}
	\label{fig:tree_structure_5d_exa}
\end{figure}

As an example we show a possible tree structure of a five-dimensional vine copula in \autoref{fig:tree_structure_5d_exa}. All appearing edges are identified with a pair-copula in the vine decomposition. For example, the third tree contains the specifications of $c_{1,3;4,5}$ and $c_{2,3;4,5}$.
%As an example we consider a five-dimensional vine copula specified by the following pair-copulas. Tree 1: $c_{1,5}$ is a Gumbel copula with $\tau_{1,5}=0.6$, $c_{2,4}$ is a BB1 copula with $\tau_{2,4}=0.83$, $c_{3,4}$ is a BB7 copula with $\tau_{3,4}=0.74$ and $c_{4,5}$ is a Tawn copula with $\tau_{4,5}=0.72$; Tree 2: $c_{1,4;5}$ is a Clayton copula with $\tau_{1,4;5}=0.5$, $c_{2,5;4}$ is a Joe copula with $\tau_{2,5;4}=0.45$ and $c_{3,5;4}$ is a BB6 copula with $\tau_{3,5;4}=0.48$; Tree 3: $c_{1,3;4,5}$ is a t copula with $\tau_{1,3;4,5}=-0.19$ and $\nu_{1,3;4,5}=3$ degrees of freedom and $c_{2,3;4,5}$ is a Frank copula with $\tau_{2,3;4,5}=-0.31$; Tree 4: $c_{1,2;3,4,5}$ is a Gaussian copula with $\tau_{1,2;3,4,5}=-0.13$. The corresponding tree structure is displayed in \autoref{fig:tree_structure_5d_exa}.

Vine copulas with arbitrary tree structure are often referred to as regular vines or in short \emph{R-vines}. Special cases of vine copula structures are so-called \textit{D-vines} and \textit{C-vines}. For a D-vine each node in tree $T_1$ has a degree of at most $2$ such that the trees are simply connected paths. In a C-vine for each tree $T_m$ there exists a root node with degree $d-m$, i.e.\ it is a neighbor of all other nodes. Each tree then has a star-like structure. 

%\cite{dissmann2013selecting} and \cite{stoeber2012} showed how the structure of a vine copula decomposition can be stored in a lower triangular matrix, a so-called vine structure matrix.

Note that in general the pair-copula $c_{j_e,k_e;D_e}$ depends on the conditioning value $\ub_{D_e}$. In order to reduce model complexity and to enable statistical inference even in high dimensions, one often makes the so-called \textit{simplifying assumption} that the influence of $\ub_{D_e}$ can be neglected and $c_{j_e,k_e;D_e}$ is equal for all possible values of $\ub_{D_e}$.

\cite{stoeber2013simplified} investigated which multivariate copulas could be represented as simplified vines: Similar to the relationship between correlation matrices and partial correlations \citep{bedford2002vines}, every Gaussian copula can be written as a simplified Gaussian vine, i.e.\ a vine copula with only bivariate Gaussian pair-copulas, where any (valid) vine structure can be used and the parameters are the corresponding partial correlations. Vice versa, every Gaussian vine represents a Gaussian copula. Further, t copulas can also be decomposed into simplified vines with arbitrary (valid) vine structure. The pair-copulas are then bivariate t copulas, the association parameters are the corresponding partial correlations and the degrees of freedom in tree $T_m$ are $\nu+(m-1)$, where $\nu$ is the degrees of freedom parameter of the t copula. However, a regular vine copula with only bivariate t copulas, called a t vine, does not necessarily represent a t copula.

The so-called \textit{Di\ss mann algorithm} \citep[cf.][]{dissmann2013selecting} is a treewise sequential algorithm that fits a simplified vine copula model to a given data set, where pairs with high dependence are modeled in lower trees in order to keep the induced estimation bias low. It is also implemented in the R package \texttt{VineCopula} \citep{VC} as \texttt{RVineStructureSelect}.

\subsection{Model distances for vine copulas}\label{sec:modeldist}
In this section we shortly review the definitions of the most important distance measures discussed in \cite{killiches2015model}. For detailed information about the concepts consult this paper and references therein. Starting point is the so-called \textit{Kullback-Leibler} \textit{distance} \citep[see][]{kullback1951information} between two $d$-dimensional copula densities $c^f$, $c^g\colon[0,1]^d\to [0,\infty)$, defined as
\begin{equation}\label{eq:KL}
\KL(c^f,c^g)=\int_{\ub\in[0,1]^d}\ln\left(\frac{c^f(\ub)}{c^g(\ub)}\right)c^f(\ub)\diff\ub.
\end{equation}

Note that due to the lack of symmetry the KL distance is not a distance in the classical sense and therefore is also referred to as \emph{Kullback-Leibler divergence}. If $c^f$ and $c^g$ are the corresponding copula densities of two $d$-dimensional densities $f$ and $g$, it can be easily shown that the KL distance between $c^f$ and $c^g$ is equal to the one between $f$ and $g$ if their marginal distributions coincide. It is common practice to use the Inference Functions for Margins (IFM) method: First the univariate margins are estimated and observation are transform to the copula scale; afterwards the copula is estimated based on the transformed data \citep[cf.][Section 10.1]{joe1997multivariate}. Therefore, it can be justified that in the remainder of the paper we restrict ourselves to data on the copula scale.

Since in the vast majority of cases the KL distance cannot be calculated analytically, the main problem of using the KL distance in practice is the computational intractability for dimensions larger than 4. There, numerical integration suffers from the curse of dimensionality and thus becomes exceptionally inefficient. As a remedy for this issue, Proposition 2 from \cite{killiches2015model} expresses the KL between multivariate densities in terms of the sum of expected KL distances between univariate conditional densities:
\begin{equation}\label{eq:KLcKL}
\KL\big(c^{f},c^{g}\big)=\sum_{j=1}^d\Ebb_{c^{f}_{(j+1):d}}\Big[\KL\Big(c^{f}_{j|(j+1):d}\left(\,\cdot\,|\Ub_{(j+1):d}\right),c^{g}_{j|(j+1):d}\left(\,\cdot\,|\Ub_{(j+1):d}\right)\Big)\Big],
\end{equation}
where for $j<d$ we use the abbreviation $(j+1)\!:\!d=\left\lbrace j+1,j+2,\ldots,d\right\rbrace $ with ${(d+1)\!:\!d}:=\emptyset$ and $\Ub_{(j+1):d}\sim c^{f}_{(j+1):d}$. It would be a valid approach to approximate the expectations in \autoref{eq:KLcKL} by Monte Carlo integration, i.e.\ the average over evaluations of the integrand on a grid of points simulated according to $c^{f}_{(j+1):d}$. Since this would also be computationally challenging in higher dimensions and additionally has the disadvantage of being random, \cite{killiches2015model} propose to approximate the expectations through evaluations on a grid consisting of only warped diagonals in the respective unit (hyper)cube. The resulting \textit{diagonal Kullback-Leibler} (dKL) distance between two $d$-dimensional R-vine models $\Rc^{f}$ and $\Rc^{g}$ is hence defined by
\[
\dKL\left(\Rc^{f},\Rc^{g}\right)=\sum_{j=1}^{d-1}\frac{1}{| \Dc_{j}|} \sum_{\ub\in \Dc_{j}} \KL\left(c^{f}_{j|(j+1):d}(\cdot|\ub),c^{g}_{j|(j+1):d}(\cdot|\ub)\right),
\]
where the set of warped discrete diagonals $\Dc_{j}\in[0,1]^{d-j}$ is given by

\[
\Dc_j=T_j\left(\left\lbrace \left\lbrace \rb+\mu \vb(\rb)\mid\mu\in\Ic_{\eps,n} \right\rbrace\, \middle|\,\rb\in\left\lbrace 0,1 \right\rbrace^{d-j}  \right\rbrace\right). 
\] 
Here, $\rb\in\left\lbrace 0,1 \right\rbrace^{d-j}$ are the corner points in the unit hypercube $\left[0,1 \right]^{d-j}$, $\vb\colon\left\lbrace 0,1 \right\rbrace^{d-j}\rightarrow\left\lbrace -1,1 \right\rbrace^{d-j}$, $\vb(\rb)=\boldsymbol{1}-2\rb$ denotes the direction vector from $\rb$ to its opposite corner point and $\Ic_{\eps,n}$ is the equidistantly discretized interval $[\eps,1-\eps]$ of length $n$. Hence $\left\lbrace \rb+\mu \vb(\rb)\mid\mu\in\Ic_{\eps,n} \right\rbrace$ represents a discretization of the diagonal from $\rb$ to its opposite corner point $\rb+\vb(\rb)$. Finally, these discretized diagonals are transformed using is the inverse Rosenblatt transformation $T_j$ with respect to $c^f_{(j+1):d}$ \citep{rosenblatt1952}. Recall that the Rosenblatt transformation $\ub$ of a vector $\wb\in[0,1]^d$ with respect to a distribution function $C$ is defined by $u_d=w_d$, $u_{d-1}=C_{d-1| d}^{-1}(w_{d-1}| u_d)$, $\ldots$ , $u_1=C_{1| 2:d}^{-1}(w_1| u_2,\ldots,u_d)$. Often it is used to transform a uniform sample on $[0,1]^d$ to a sample from $C$. The concept is used to transform the unit hypercube's diagonal points to points with high density values of $c^f_{(j+1):d}$
%Here, $T_j$ is the inverse Rosenblatt transformation \citep{rosenblatt1952} with respect to $c^f_{(j+1):d}$ and the equidistantly discretized interval $[\eps,1-\eps]$ of length $n$ is denoted by $\Ic_{\eps,n}$. Further, the direction function $\vb(\cdot)$ is given by $\vb\colon\left\lbrace 0,1 \right\rbrace^{d-j}\rightarrow\left\lbrace -1,1 \right\rbrace^{d-j}$, $\vb(\rb)=\boldsymbol{1}-2\rb$. 
Hence, the KL distance between $c^f$ and $c^g$ is approximated by evaluating the KL distances between the univariate conditional densities $c^{f}_{j|(j+1):d}$ and $c^{g}_{j|(j+1):d}$ conditioned on values lying on warped diagonals $\Dc_j$, $j=1,\ldots,d-1$. Diagonals have the advantage that all components take values on the whole range from 0 to 1 covering especially the tails, where the substantial differences between copula models occur most often. With the above modifications the intractability of the KL for multivariate densities is overcome since only KL distances between univariate densities have to be evaluated. It was shown in Proposition 1 of \cite{killiches2015model} that these univariate conditional densities $c^{f}_{j|(j+1):d}$ and $c^{g}_{j|(j+1):d}$ can be easily derived for the vine copula model. Moreover, in {Remark 1} they prove that for $\eps\to 0$ and $n\to\infty$ the dKL converges to a sum of scaled line integrals. Further, they found heuristically that even for $n=10$ and $\eps=0.025$ the dKL was a good and fast substitute for the KL distance. 

Nevertheless, since the number of diagonals grows exponentially in the dimension, \cite{killiches2015model} found that for really high dimensions (e.g.\ $d>20$) the computation of the dKL was rather slow. However, they illustrated in several examples that the restriction of the evaluations to a single \textit{principle diagonal} yields exceptionally fast calculations and still results in a viable distance measure with qualitative outcomes close to the KL distance. For $j<d$, out of the set $\mathbb{D}_j$ of the $2^{(d-j)-1}$ possible $(d-j)$-dimensional diagonals, the principal diagonal $\Dc_j^*$ is the one with the largest weight, measured by an integral over the diagonal with respect to the density $c^f_{(j+1):d}$, i.e.\ \[\Dc_j^*=\argmax\limits_{\Dc\in\mathbb{D}_j}\int_{\xb\in D}c^f_{(j+1):d}(\xb)\diff\xb.\] 
Hence, the \textit{single diagonal Kullback-Leibler distance} between $\Rc^{f}$ and $\Rc^{g}$ is given by
\[
\sdKL(\Rc^{f},\Rc^{g})=\sum_{j=1}^{d-1} \frac{1}{| \Dc^*_{j}|} \sum_{\ub\in \Dc^*_{j}} \KL\left(c^{f}_{j|(j+1):d}(\cdot|\ub),c^{g}_{j|(j+1):d}(\cdot|\ub)\right).
\]

In practice it has shown to be advisable to use the dKL for dimensions $d<10$ and the sdKL for higher-dimensional applications. For a more detailed discussion on implementation and performance of these distance measures we refer to \cite{killiches2015model}.

Without further reference it is not possible to decide whether a given distance measure value is large or small. Therefore in the following we develop a statistical test to decide whether a distance value is significantly different from zero.

\subsection{Hypothesis test for model selection}\label{sec:paramboots}

\begin{figure}[!htbp]
	\centering
	\includegraphics[width=\textwidth]{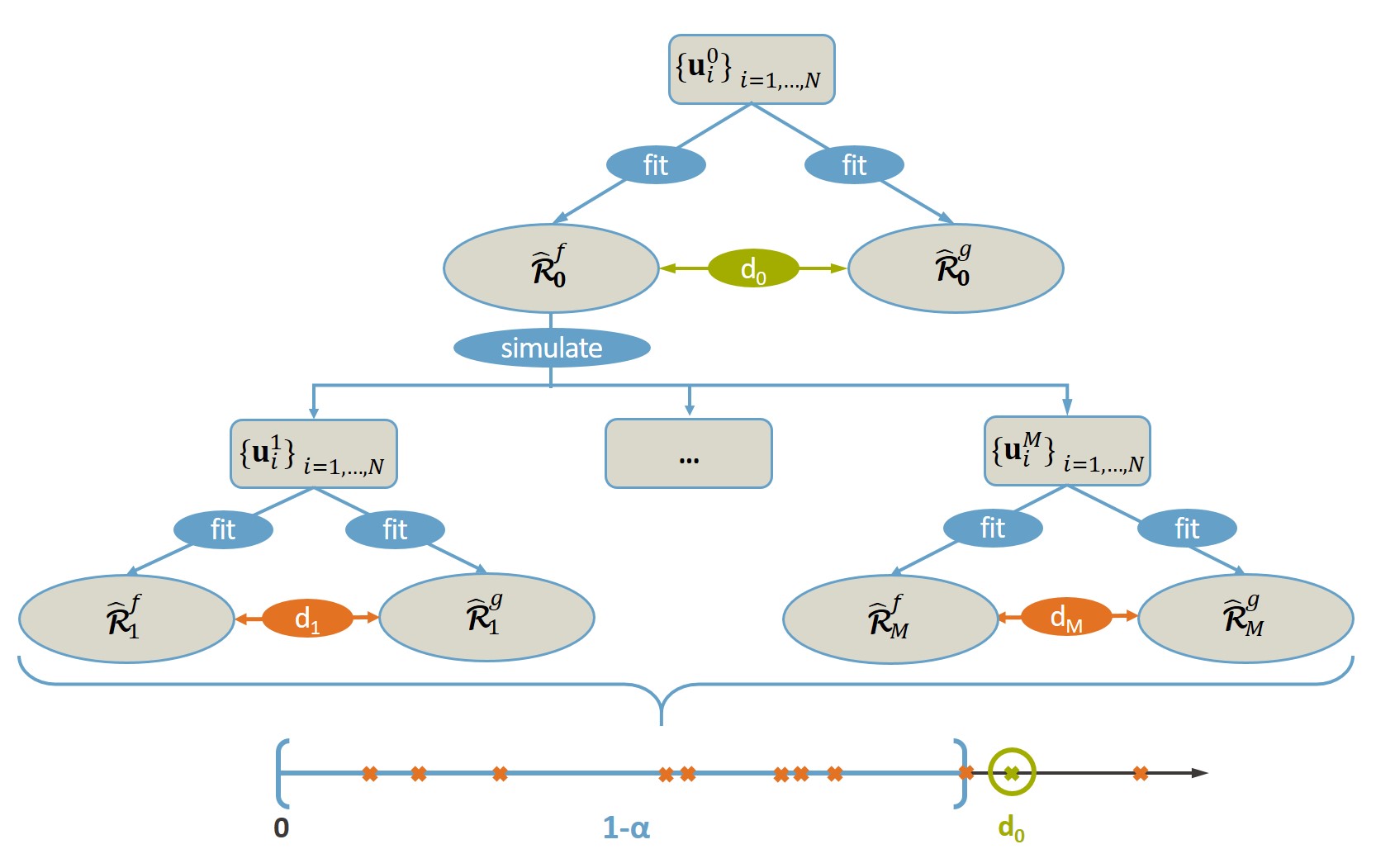}
	\caption{Scheme of the testing procedure based on parametric bootstrapping.}
	\label{fig:test_scheme}
\end{figure}

In this section we provide a procedure based on parametric bootstrapping \citep[see][]{efron1994introduction} for choosing between a parsimonious and a more complex model. Assume we have two nested classes of $d$-dimensional parametric copula models $\mathbb{C}^f\subseteq\mathbb{C}^g$ and a copula data set $\ub_i^0\in [0,1]^{d}$, $i=1,\ldots,N$, with true underlying distribution $\Rc^g\in\mathbb{C}^g$. We want to investigate whether a model from $\mathbb{C}^f$ suffices to describe the data. In other words, we want to test the null hypothesis $H_0\colon \Rc^g\in \mathbb{C}^f$, which means that there exists $\Rc^f\in\mathbb{C}^f$ such that $\Rc^f=\Rc^g$. Due to the identity of indiscernibles (i.e.\ $\KL(\Rc^f,\Rc^g)=0$ if and only if $\Rc^f=\Rc^g$) of the Kullback-Leibler distance this is equivalent to $\KL(\Rc^f,\Rc^g)=0$. Hence, for testing $H_0$ we can examine whether the KL distance between $\Rc^f$ and $\Rc^g$ is significantly different from zero. In practice, $\Rc^f$ and $\Rc^g$ are unknown and have to be estimated from the data $\ub_i^0\in [0,1]^{d}$. Consider the KL distance $d_0=\KL(\hat{\Rc}_0^f, \hat{\Rc}_0^g)$ between the two fitted models $\hat{\Rc}_0^f$ and $\hat{\Rc}_0^g$ as the test statistic. Since the distribution of $d_0$ cannot be derived analytically we use the following parametric bootstrapping scheme to retrieve it:

%Further, let $\Rc^{f}\in\mathbb{C}^0$ and $\Rc^{g}\in\mathbb{C}^1$ be two fitted copulas. In order to assess if the more parsimonious $\Rc^f$ suffices to describe the data properly, we test whether the distance $d_0$ between $\Rc^{f}$ and $\Rc^{g}$ is significantly different from zero, with respect to some distance measure $\Delta\colon\mathbb{C}^0\times\mathbb{C}^1\to[0,\infty)$. This is equivalent to testing the null hypothesis $H_0\colon \Rc^*\in\mathbb{C}^0$. In order to test $H_0$ we perform the following test with significance level $\alpha$. First calculate the distance $d_0$ between them using the distance measure $\Delta$:
%\[
%d_0=\Delta(\Rc^{f},\Rc^{g}).
%\]
%Then, we repeat the following steps 
For $j=1,\ldots,M$, generate a sample $\ub^j_i\in [0,1]^{d}$, $i=1,\ldots,N$, from $\hat{\Rc}_0^f$. Fit copulas $\hat{\Rc}^f_{j}\in\mathbb{C}^f$ and $\hat{\Rc}^g_{j}\in\mathbb{C}^g$ to the generated sample. Calculate the distance between $\hat{\Rc}^f_{j}$ and $\hat{\Rc}^g_{j}$:
\[
d_j=\KL (\hat{\Rc}^f_{j}, \hat{\Rc}^g_{j}).
\]
Now reorder the set $\left\lbrace d_j \mid j=1,\ldots, M\right\rbrace$ such that $d_1<d_2<\ldots<d_M$. For a significance level $\alpha\in(0,1)$, we can determine an empirical confidence interval $I_{1-\alpha}^M\subseteq \left[0,\infty\right)$ with confidence level $1-\alpha$ by 
\[
I_{1-\alpha}^M=[0,d_{\lceil M(1-\alpha) \rceil}],
\]
where $\lceil \cdot\rceil$ denotes the ceiling function. Finally, we can reject $H_0$ if $d_0\notin I_{1-\alpha}^M$. \autoref{fig:test_scheme} illustrates the above procedure in a flow chart. At the bottom the resulting distances are plotted on the positive real line. In this exemplary case, $d_0$ (circled cross) lies outside the range of the empirical $100(1-\alpha)\%$ confidence interval and therefore $H_0$ can be rejected at the $100\alpha\%$ level, i.e.\ there is a significant difference between $\hat{\Rc}_0^f$ and $\hat{\Rc}_0^g$.		

Since in higher dimensions the KL distance cannot be calculated in a reasonable amount of time, we use the distance measures dKL (for $d<10$) and sdKL (for $d\geq 10$), introduced in \autoref{sec:modeldist}, as substitutes for the KL distance. The above bootstrapping scheme works similarly using the substitutes.

Of course, it is not obvious per se how to choose the number of bootstrap samples $M$. On the one hand we want to choose $M$ as small as possible (due to computational time); on the other hand we want the estimate of $d_{\lceil M(1-\alpha) \rceil}$ to be as precise as possible in order to avoid false decisions with respect to the null hypothesis (the upper bound $d_{\lceil M(1-\alpha) \rceil}$ of the confidence interval $I_{1-\alpha}^M$ is random with variance decreasing in $M$). Therefore, we choose $M$ so large that $d_0$ lies outside the $100(1-\beta)\%$ confidence interval of $d_{\lceil M(1-\alpha) \rceil}$ such that we can decide whether $d_0$ is significantly larger (smaller) than  $d_{\lceil M(1-\alpha) \rceil}$. This confidence interval can be obtained from an estimate of the distribution of the $\lceil M(1-\alpha) \rceil$th order statistic \citep[see for example][page 232]{casella2002statistical}. In all applications of this test contained in this paper we found that for $\alpha=5\%$ and $\beta=1\%$ a bootstrap sample size of $M=100$ was enough.

%Every time we use the test, we generate an amount of samples that ensures that the histogram of the obtained distances $d_j$, $j=1,\ldots,M$, is smooth and $d_0$ lies either clearly inside or outside the empirical $100(1-\alpha)\%$ confidence interval. This is sufficient for our purpose since we are only interested in an answer to the question whether the difference between two models is significant and not in the exact p-value. We found that in our applications a bootstrap sample of $M=100$ was already enough.

\paragraph{Validity of the parametric bootstrap for the hypothesis test}\mbox{}\\
In order to establish the validity the parametric bootstrap for the above hypothesis test we will argue that under the null hypothesis $H_0$ the bootstrapped distances $d_j$, $j=1,\ldots,M$, are i.i.d.\ with a common distribution that is close to $F_{d_0}$, i.e.\ the distribution of $d_0$, for large sample size $N$: If we have a consistent estimator for $\Rc^f$, we know that under $H_0$ the estimate $\hat\Rc^f$ is close to $\Rc^f$ for large $N$. Since the bootstrap samples $\ub_i^j$, $i=1,\ldots,N$, $j=1,\ldots,M$ are generated from $\hat\Rc^f$, they can be assumed to be approximate samples from $\Rc^f$. Since $\hat\Rc^f_j$ and $\hat\Rc^g_j$ are estimated based on the $j$th bootstrap sample $\ub_i^j$, $i=1,\ldots,N$, the KL between $\hat\Rc^f_j$ and $\hat\Rc^g_j$, i.e.\ $d_j$, has the same distribution as the KL between $\hat\Rc^f$ and $\hat\Rc^g$, i.e.\ $d_0$, for large $N$. Therefore, we can construct empirical confidence intervals for $d_0$ based on the bootstrapped distances $d_j$, $j=1,\ldots,M$. 

Of course this argumentation is not a strict proof but rather makes the proposed approach plausible. An example for a mathematical justification of the parametric bootstrap in the copula context can be found in \cite{genest2008validity}. In \autoref{sec:power} we will see in a simulation study that our proposed test holds its level under the null hypothesis (for different sample sizes) when investigating the power of the test in a simplified/non-simplified vine copula framework.

%Since $\Rc^f$ and $\Rc^g$ are assumed to be parametric models 
% consistent estimate $\hat\Rc^f$ of $\Rc^f$
%
%consider the asymptotic behavior of the test statistic $d_0$ under the null hypothesis $H_0$. For this let $\hat\Rc^f(N)$ and $\hat\Rc^g(N)$ be the parametric estimates of $\Rc^f$ and $\Rc^f$, respectively, based on a sample $\ub_i^0\in[0,1]^d$, $i=1,\ldots,N$, of size $N$. Further, denote by $d_0^\infty$ the limit of the test statistic $d_0(N)=\KL(\hat{\Rc}_0^f(N), \hat{\Rc}_0^g(N))$.
%
%
%
%Under $H_0$ we have that $\hat\Rc^f_n\to \Rc^f$ for $n\to\infty$ (for a consistent estimator) and 
%all model depnd on parameters
%can be estimated consistently
%therefore, asympotically correct models
%although we do not derive 
%
%Regarding the validity of the parametric bootstrap for the above hypothesis test \cite{genest2008validity} .
%
%Verweis auf Paper genest/remi \\
%laueft bei uns analog, weil modelle nur von parametern abhaengen. 

\section{Testing simplified versus non-simplified vine copulas}\label{sec:simplified}

As already mentioned in the introduction, the validity of the simplifying assumption is a frequently discussed topic in the recent literature. For the case the simplifying assumption is not satisfied, \cite{vatter2015gamvine} developed a method to fit a non-simplified vine to given data such that the parameters of the pair-copulas with non-empty conditioning sets are dependent on the conditioning variable(s). This functional relationship is modeled with a generalized additive model. The fitting algorithm is implemented in the R package \texttt{gamCopula} \citep{vatter2015gamCopula} as the function \texttt{gamVineStructureSelect}. The selection of the vine structure is identical to the one of \texttt{RVineStructureSelect}.

In this section we will present how distance measures can be used to decide whether a (more complicated) non-simplified model is needed or the simplified model suffices. This can be done with the help of the test introduced in \autoref{sec:paramboots} (using the dKL). Here we take $\mathbb{C}^f$ and $\mathbb{C}^g$ to be the class of simplified and non-simplified vine copula models, respectively. Since every simplified vine can be represented as a non-simplified vine with constant parameters, $\mathbb{C}^f$ and $\mathbb{C}^g$ are nested, i.e.\ $\mathbb{C}^f\subseteq\mathbb{C}^g$. Now, the null hypothesis $H_0$ to be tested at significance level $\alpha$ is that the true underlying model is in $\mathbb{C}^f$, i.e.\ the model is simplified. 

After investigating the power of the test (\autoref{sec:power}) we apply it to a hydro-geochemical and a financial data set in \autoref{sec:satestExamples}.

\subsection{Power of the test}\label{sec:power}
In a simulation study we investigate the performance of our test. For this purpose we consider a three-dimensional non-simplified vine consisting of the pair-copulas $c_{1,2}$, $c_{2,3}$ and $c_{1,3;2}$, where all pairs are bivariate Clayton copulas. The Kendall's $\tau$ values of the copulas $c_{1,2}$ and $c_{2,3}$ are $\tau_{1,2}=0.7$ and $\tau_{2,3}=0.5$, respectively. The third $\tau$ value depends linearly on $u_2$: $\tau_{1,3;2}(u_2)=a+(b-a) u_2$ with constants $a,b\in[-1,1]$. For $a=b$ the function is constant such that the vine is simplified. By construction $\tau_{1,3;2}$ can become negative for some combinations of $a$, $b$ and $u_2$; in such cases we use the 90 degree rotated version of the Clayton copula since the Clayton copula does not allow for negative dependence.

\begin{figure}[!htbp]
	\centering
	\includegraphics[trim=1mm 6mm 5mm 17mm, clip,width=0.7\textwidth]{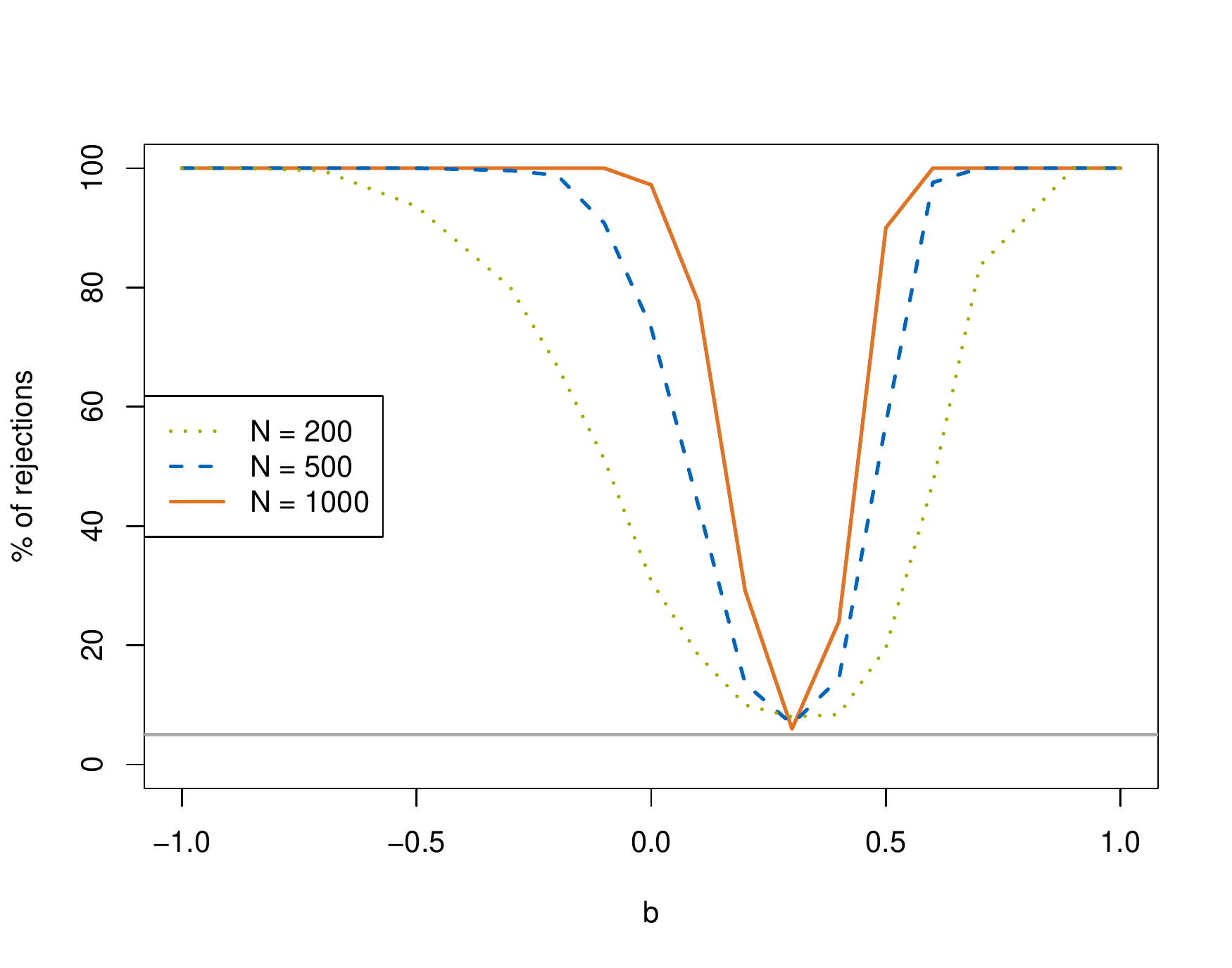}
	\caption{Percentage of rejections of $H_0$ (at level $\alpha=5\%$) depending on $b$ with constants $a=0.3$, $M=100$ and $P=250$ for $N=200$ (dotted curve), $N=500$ (dashed curve) and $N=1000$ (solid curve). The horizontal solid gray line indicates the 5\% level.}
	\label{fig:PowerTest}
\end{figure}

By fixing $a=0.3$ and letting $b$ range between $-1$ and $1$ in $0.1$ steps we obtain 21 scenarios. For each of the scenarios we generate a sample of size $N\in \left\{200, 500, 1000\right\}$ from the corresponding non-simplified vine copula and fit both a simplified and a non-simplified model to the generated data. Since we are only interested in the parameters and their variability we fix both the vine structure and the pair-copula families to the true ones. We test the null hypothesis that the two underlying models are equal. In order to assess the power of the test, we perform this procedure $P=250$ times (at significance level $\alpha=5\%$) and check how many times the null hypothesis is rejected. As sample size we take the same $N$ used to generate the original data. In each test we perform $M=100$ bootstrap replications. In \autoref{fig:PowerTest} the proportions of rejections of the null hypothesis within the $P=250$ performed tests are shown depending on $b$. The different sample sizes are indicated by the three different curves: $N=200$ (dotted curve), $N=500$ (dashed curve) and $N=1000$ (solid curve).

%\begin{table}[!htb]
%	\centering
%	\begin{tabular}{|c|cccc|}
%		\hline
%		$b$ & -$0.7$ & $0.1$ & $0.3$ & $0.9$\\
%		\hline
%		$\%$ of rejections & $97.6\%$ & $82.8\%$ & $5.2\%$ & $100\%$\\
%		\hline
%	\end{tabular}
%	\caption{Percentage of rejections of $H_0$ for different values of $b$ with constants $a=0.3$, $n=200$, $M=100$ and $P=250$.}
%	\label{tab:power}
%\end{table}\\

%% MATTHIAS
We see that the observed power of the test is in general very high. Considering the dashed curve, corresponding to a sample size of $N=500$, one can see the following: If the distance $\left| b-a\right| $ is large, $\tau_{1,3;2}$ is far from being constant. Hence, we expect the non-simplified vine and the simplified vine to be very different and therefore the power of the test to be large. We see that for $b\leq -0.1$ and $b\geq 0.6$ the power of the test is above $80\%$ and for $b\leq -0.2$ and $b\geq 0.7$ it is even (close to) $100\%$. For values of $b$ closer to $a$ the power decreases. For example, for $b=0.1$ the Kendall's $\tau$ value $\tau_{1,3;2}$ only ranges between 0.1 and 0.3 implying that the non-simplified vine does not differ too much from a simplified vine. Therefore, we cannot expect the test to always detect this difference. Nevertheless, even in this case the power of the test is estimated to be almost $44\%$. From a practical point of view, this result is in fact desirable since models estimated based on real data will always exhibit at least slight non-simplifiedness due to randomness even when the simplifying assumption is actually satisfied. Further, for $b=0.3$ the function $\tau_{1,3;2}(u_2)$ is actually constant with respect to $u_2$ so that $\Rc^*$ is a simplified vine. Thus, $H_0$ is true and we hope to be close to the significance level $\alpha=5\%$. With $6.4\%$ of rejections, we see that this is the case here.

Looking at the dotted and the solid curve we find that the higher the sample size is, the higher is the power of the test, which is what one also would have expected.  In the case of $N=1000$, we have a power of over $80\%$ for $b\in[-1,0]\cup [0.5,1]$ and even $100\%$ rejections for $b\in[-1,-0.1]\cup [0.6,1]$. For $b=0.3$ the test holds its level with $5.2\%$ of rejections. Yet even for a sample size of as little as $N=200$, the power of the test is above $80\%$ for values of $b$ between $-1$ and $-0.3$ and $0.5$ and $1$. A power of $100\%$ is reached for $b\leq -0.8$ and $b\geq 0.9$. For $b=0.3$ the test rejects the null hypothesis in $7.2\%$ of the cases.

We can conclude that our test is a valid $\alpha$-level method in finite samples to decide if a non-simplified model is necessary.

\subsection{Real data examples}\label{sec:satestExamples}
\paragraph{Three-dimensional subset of the uranium data set}
To show an application of our test we use a subset of the classical seven-dimensional hydro-geochemical data set \citep{cook1986generalized}, which has amongst others been investigated by \cite{acar2012beyond} and \cite{killiches2016examination} with respect to the simplifying assumption. The data set consists of $N=655$ observations of log concentrations of different chemicals in water samples from a river near Grand Junction, Colorado. We will focus on the three chemicals cobalt ($U_1$), titanium ($U_2$) and scandium ($U_3$) and fit both a simplified and a non-simplified vine copula to the data.

\begin{figure}[!htbp]
	\centering
	\includegraphics[trim=0.6cm 1.1cm 0.9cm 2.0cm, clip,width=0.4\textwidth]{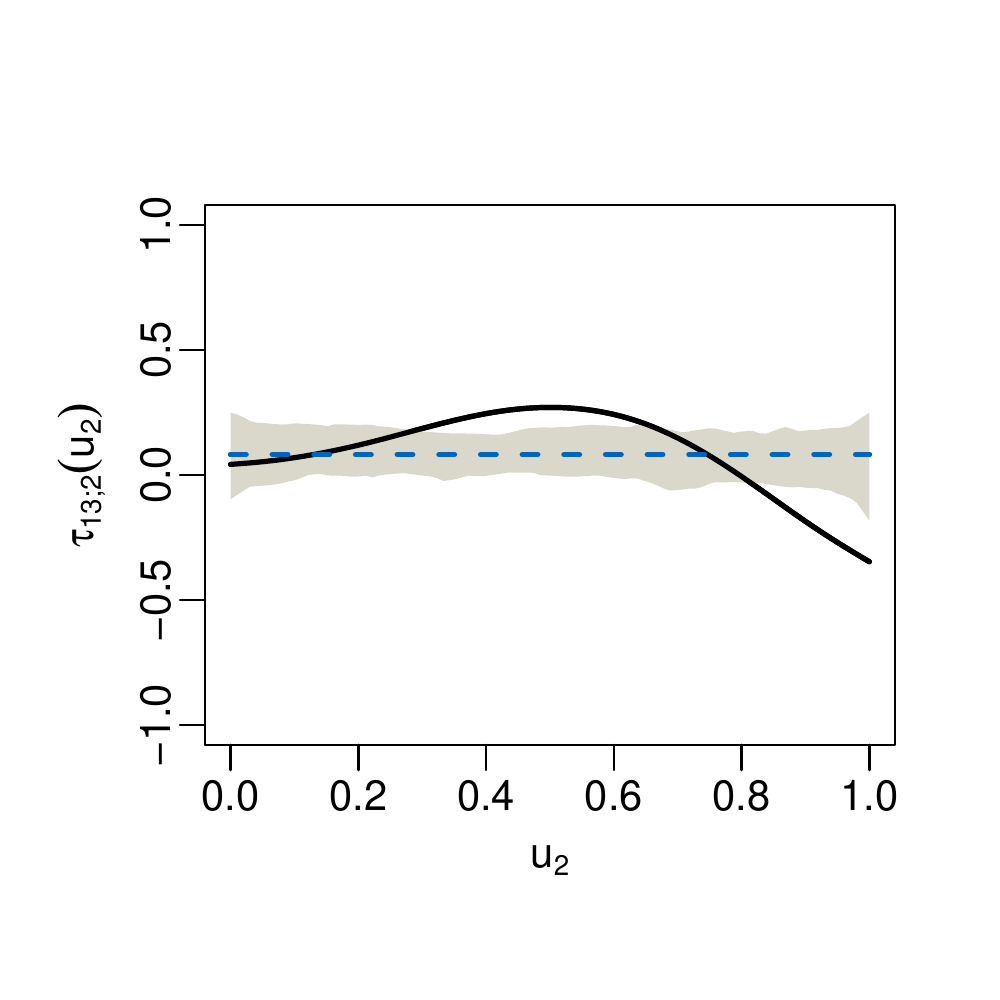}
	\caption{Estimated functional relationship between $\tau_{1,3;2}$ and $u_2$ (for the non-simplified model $\hat\Rc_0^g$). The dashed line represents the constant $\tau_{1,3;2}$ of the simplified model $\hat\Rc_0^f$ and the gray area indicates the pointwise bootstrapped 95\% confidence bounds under $H_0$.}
	\label{fig:tau13_2}
\end{figure}

The fitted simplified vine $\hat\Rc_0^f$ is specified in the following way: $c_{1,2}$ is a t copula with $\tau_{1,2}=0.53$ and $\nu_{1,2}=8.03$, $c_{2,3}$ is a t copula with $\tau_{2,3}=0.43$ and $\nu_{2,3}=5.93$ and $c_{1,3;2}$ is a t copula with $\tau_{1,3;2}=0.08$ and $\nu_{1,3;2}=5.65$. For the non-simplified vine $\hat\Rc_0^g$, the pair-copulas $c_{1,2}$ and $c_{2,3}$ are the same as for the simplified vine, $c_{1,3;2}$ is also still a t copula but now has $\nu_{1,3;2}=6.69$ degrees of freedom and its association parameter depends on $u_2$ as displayed as the solid line in \autoref{fig:tau13_2}. For values of $u_2$ below 0.8 (roughly) we have small positive Kendall's $\tau$ values, whereas for the remaining values we observe small to medium negative association. For comparison, the (constant) Kendall's $\tau$ of the estimated simplified vine is plotted as a dashed line. Further, the pointwise bootstrapped 95\% confidence bounds under $H_0$ are indicated by the gray area.

The fact that the estimated Kendall's $\tau$ function exceeds these bounds for more than half of the $u_2$ values suggests that the simplified and non-simplified vines are significantly different from each other. We now use our testing procedure to formally test this.

The distance between the two vines is $\dKL_0=0.058$. In order to test the null hypothesis we generate $M=100$ samples of size $N=655$ from $\hat\Rc_0^f$. Then, for each sample we estimate a non-simplified vine $\hat\Rc^{g}_{j}$ and a simplified vine $\hat\Rc^{f}_{j}$ and calculate the distance between them.
%In \autoref{fig:boxploturanium} the resulting 100 dKL-values are displayed in a boxplot.
%\begin{figure}[!htb]
%    \centering
% \includegraphics[width=\textwidth]{plots/uranium_ParBootstrap_BP_rotated.pdf}
%    \caption{Boxplot of the distance values $\dKL_r$ and $\dKL_0$ (filled diamond shaped point).}
%    \label{fig:boxploturanium}
%\end{figure}\\
%The filled diamond shaped point to the right corresponds to the distance between $\Rc^0$ and $\Rc^1$.
Since with one exception all resulting simulated distances are considerably smaller than $\dKL_0$, we can reject $H_0$ at the 5\% level (with a p-value of 0.01). Hence, we conclude that here it is necessary to model the dependence structure between the three variables using a non-simplified vine. \cite{acar2012beyond} and \cite{killiches2016examination} also come to the conclusion that a simplified vine would not be sufficient in this example.

%To compare our results we also repeat the parametric bootstrap procedure with the MCKL as a distance measure. We use the $M=100$ fitted non-simplified vines $\Rc^{1,r}$ from above and estimate the KL-distance to $\Rc^{0}$ by Monte Carlo with $N_{MC}=10{,}000$. The results (see \autoref{fig:boxploturanium}, lower panel) look fairly similar to the ones we obtained from the sdKL.
%
%The distance $\MCKL_0$ between $\Rc^{1}$ and $\Rc^{0}$ is considerably larger than all distance values obtained from the bootstrapped vines. Thus, we would also conclude that the null hypothesis can be rejected, i.e.\ the dependence structure of the data cannot be properly described by a simplified vine. However, an average MCKL calculation took about 98.1 seconds, which is almost six times the average duration of the corresponding sdKL calculations. Further, we even observe negative MCKL-values in a few cases although the actual value of the integral to be estimated is non-negative. This again shows that the results of the sdKL are sensible and in line with other comparable methods but outperform those especially regarding computational time.

\paragraph{Four-dimensional subset of the EuroStoxx 50 data set}
We examine a 52-dimensional EuroStoxx50 data set containing 985 observations of returns of the EuroStoxx50 index, five national indices and the stocks of the 46 companies that were in the EuroStoxx50 for the whole observation period (May 22, 2006 to April 29, 2010). This data set will be studied more thoroughly in \autoref{sec:trunc_real_data}. Since fitting non-simplified vine copulas in high dimensions would be too computationally demanding we consider only a four-dimensional subset containing the following national indices: the German DAX ($U_1$), the Italian MIB ($U_2$), the Dutch AEX ($U_3$) and the Spain IBEX ($U_4$) (see also Example 1 of \cite{killiches2015model}, where this data set was already investigated). In practice it is very common to model financial returns using multivariate t copulas \citep[see e.g.][]{demarta2005t}. From \cite{stoeber2013simplified} we know that any multivariate t copula can be represented as a vine satisfying the simplifying assumption. With our test we can check whether this necessary condition is indeed fulfilled for this particular financial return data set.

\begin{figure}[!htbp]
	\centering
	\includegraphics[trim=0.6cm 1.1cm 0.9cm 2.0cm, clip,width=0.32\textwidth]{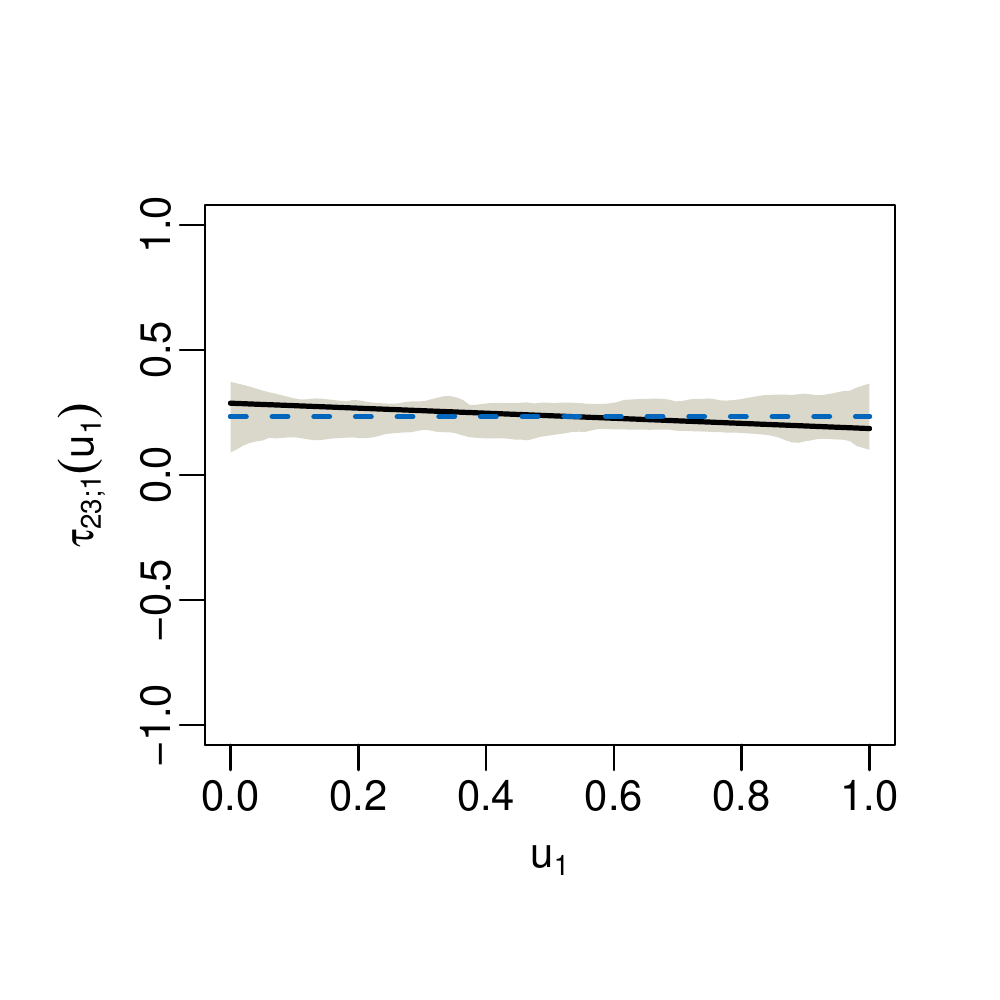}
	\includegraphics[trim=0.6cm 1.1cm 0.9cm 2.0cm, clip,width=0.32\textwidth]{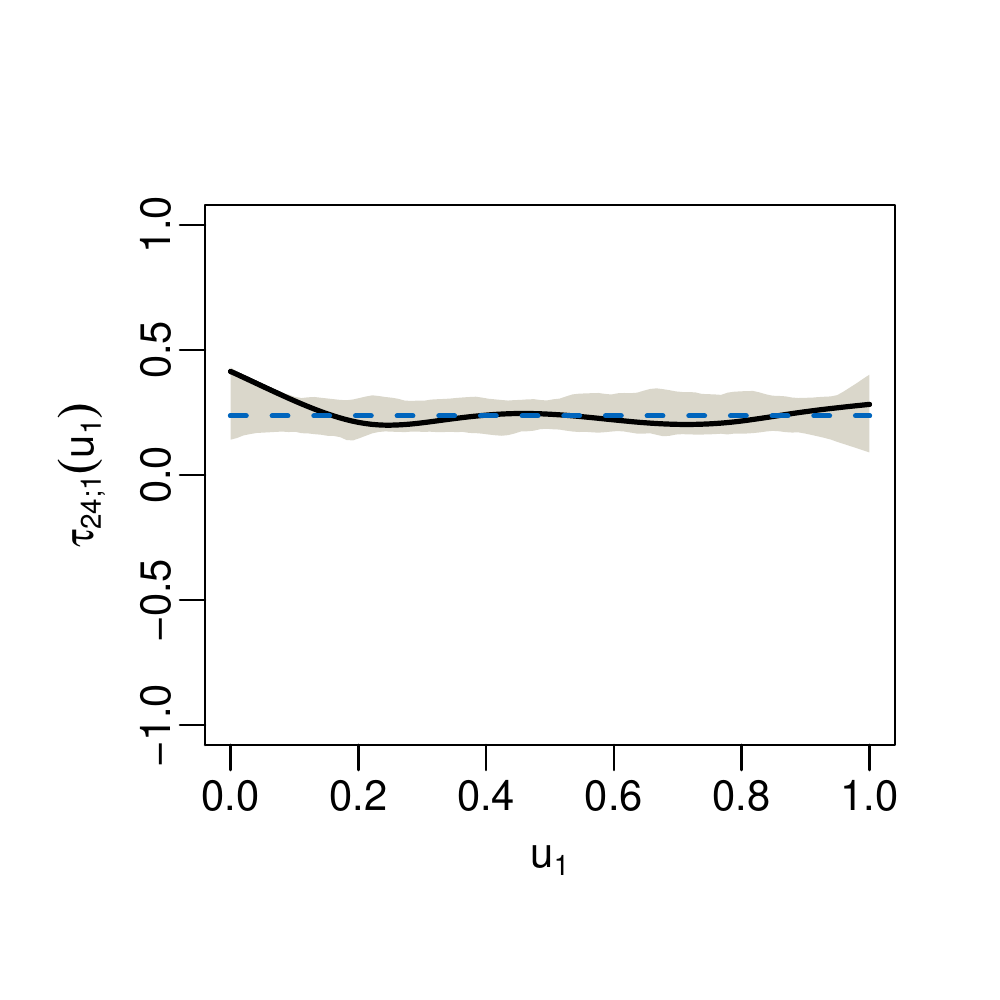}
	\includegraphics[trim=0.6cm 1.1cm 0.9cm 2.0cm, clip,width=0.32\textwidth]{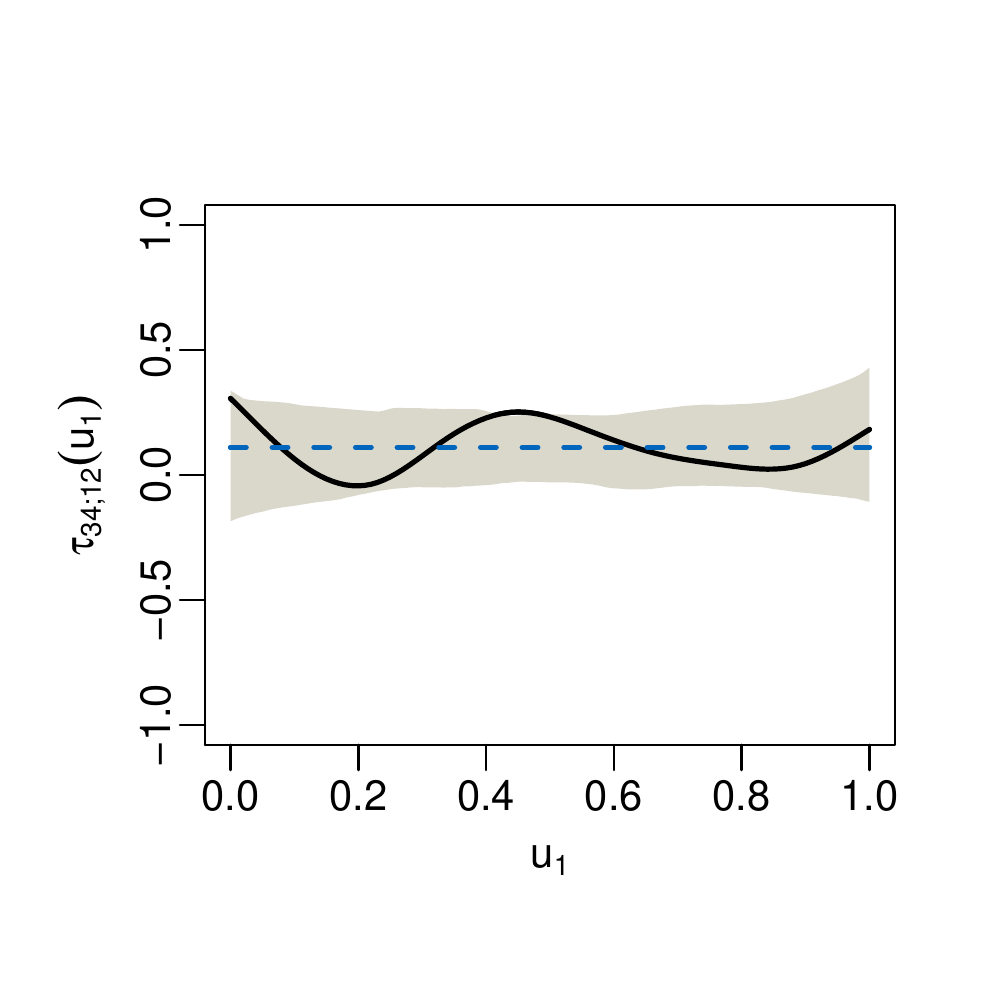}
	\caption{Estimated functional relationship of $\tau_{2,3;1}$ (left), $\tau_{2,4;1}$ (middle) and $\tau_{3,4;1,2}$ (right) in terms of $u_1$ from $\hat\Rc_0^g$.  The dashed lines represent the constant $\tau$ values of the simplified model $\hat\Rc_0^f$ and the gray areas indicate the pointwise bootstrapped 95\% confidence bounds under $H_0$.}
	\label{fig:ES50_tauplots}
\end{figure}

We proceed as in the previous section and fit a simplified model $\hat\Rc_0^f$ as well as a non-simplified model $\hat\Rc_0^g$ to the data. The estimated structures of both models are C-vines with root nodes DAX, MIB, AEX and IBEX. Again, the pair-copulas in the first tree coincide for both models being fitted as bivariate t copulas with $\tau_{1,2}=0.70$ and $\nu_{1,2}=4.96$, $\tau_{1,3}=0.72$ and $\nu_{1,3}=6.23$, and $\tau_{1,4}=0.69$ and $\nu_{1,2}=6.80$. In the second tree of the simplified model the pair-copulas are also estimated to be t copulas with $\tau_{2,3;1}= 0.23$ and $\nu_{1,2}=6.34$, and $\tau_{2,4;1}= 0.24$ and $\nu_{1,2}=10.77$. The corresponding non-simplified counterparts fitted by the \texttt{gamVineStructureSelect} algorithm are also t copulas, whose strength of dependence varies only very little and stays within the confidence bounds of the simplified vine (see \autoref{fig:ES50_tauplots}, left and middle panel). The estimated degrees of freedom are also quite close to the simplified ones ($\nu_{2,3;1}=6.47$ and $\nu_{2,4;1}=11.56$), such that regarding the second tree we would presume that the distance between both models is negligible. Considering the copula $c_{3,4;1,2}$ in the third tree, the simplified fit is a Frank copula with $\tau_{3,4;1,2}=0.11$, while the non-simplified fit is a Gaussian copula whose $\tau$ values only depend on $u_1$ (i.e.\ the value of the DAX). In the right panel of \autoref{fig:ES50_tauplots} we see the estimated relationship, which is a bit more varying than the others but still mostly stays in between the confidence bounds. The broader confidence bounds can be explained by the increased parameter uncertainty for higher order trees of vine copulas arising due to the sequential fitting procedure.

The question is now, whether the estimated non-simplified vine is significantly different from the simplified one, or in other words: Is it necessary to use a non-simplified vine copula model for this data set or does a simplified one suffice?

In order to answer this question we make use of our test using parametric bootstrapping and produce $M=100$ simulated distances under the null hypothesis that both underlying models are equivalent.
%These are visualized in form of a boxplot in \autoref{fig:boxplotES50}.
%\begin{figure}[!htb]
%    \centering
% \includegraphics[width=\textwidth]{plots/ES50_4dim_ParBootstrap_BP_rotated.pdf}
%    \caption{Boxplot of the distance values $\dKL_r$ and $\dKL_0$ (diamond shaped point).}
%    \label{fig:boxplotES50}
%\end{figure}
%% MATTHIAS
%Again, the original distance between $\Rc_0$ and $\Rc_1$ is marked as a diamond shaped point.
In this case the original distance between $\hat\Rc_0^f$ and $\hat\Rc_0^g$ is close to the lower quartile and therefore the null hypothesis clearly cannot be rejected. So we can conclude that for this four-dimensional financial return data set a simplified vine suffices to reasonably capture the dependence pattern.

In a next step we test with the procedure from \autoref{sec:paramboots} ($\alpha=5\%$, $M=100$) whether there is a significant difference between the above fitted simplified vine and a t copula. With a p-value of $0.10$ we cannot reject the null hypothesis that the two underlying models coincide such that it would be justifiable to assume a t copula to be the underlying dependence structure of this financial return dataset.

Although we only presented applications in dimensions 3 and 4, in general the procedure can be used in arbitrary dimensions. The computationally limiting factor is the fitting routine of the non-simplified vine copula model, which can easily be applied up to 15 dimensions in a reasonable amount of time \citep[for the methods implemented in ][]{vatter2015gamCopula}.

\section{Model selection}\label{sec:model_selection}
A typical application of model distance measures is model selection. Given a certain data set one often has to choose between several models with different complexity and features. Distance measures are a convenient tool that can help with the decision for the ``best'' or ``most suitable'' model out of a set of candidate models.

%%% http://staff.ustc.edu.cn/~cgong821/Wiley.Interscience.Elements.of.Information.Theory.Jul.2006.eBook-DDU.pdf

\subsection{KL based model selection}\label{sec:GOF}
The Kullback-Leibler distance is of particular interest for model selection because of the following relationship: For given copula data $\ub_i\in[0,1]^d$, $i=1,\ldots,N$, from a $d$-dimensional copula model $c\colon [0,1]^d\to [0,\infty)$ we have
\[
\KL(c,c^\perp)\approx\sum_{i=1}^N \log\left(\frac{c(\ub_i)}{c^\perp(\ub_i)}\right)=\sum_{i=1}^N \log\left(c(\ub_i)\right)=\log \ell(c),
\]
where $c^\perp$ denotes the density of the $d$-dimensional independence copula and $\log \ell(c)$ is the log-likelihood of the model $c$. This means that the log-likelihood of a model can be approximated by calculating its Kullback-Leibler distance from the corresponding independence model \citep[also known as \emph{mutual information} in the bivariate case; see e.g.][]{cover2012elements}. The log-likelihood itself as well as the information criteria AIC and BIC \citep{akaike1998information,schwarz1978estimating}, which are based on the log-likelihood but penalize the use of too many parameters, can be used to assess how well a certain model fits the data. The higher (lower) the log-likelihood (AIC/BIC) is, the better the model fit. Thus, a high Kullback-Leibler distance from the independence copula also corresponds to a good model fit. Note that this approximation only holds if data in fact was generated from $c$. Since in applications the true underlying distribution $c$ is unknown, we can use the KL distance between a fitted copula and the independence copula as a proxy for the quality of the fit. Therefore, having fitted different models to a data set it is advisable to choose the one with the largest KL distance. Since dKL and sdKL are modifications of the original KL distance, it is natural to use them as substitutes for the model selection procedure. 

In the following subsections we provide two examples, where dKL and sdKL based measures are applied for model selection. For this purpose we perform the following procedure 100 times: We fix a vine copula model and generate a sample of size $N=3000$ from it. Then, we fit different models and calculate the distance to the independence model with respect to dKL and sdKL, respectively. The results are compared to AIC and BIC.

\subsection{Five-dimensional mixed vine}\label{sec:5dmixed}
As a first example we consider a five-dimensional vine copula with the vine tree structure given in \autoref{fig:tree_structure_5d_exa} from \autoref{sec:vinecopulas} and the following pair-copulas:
\begin{itemize}
	 \item Tree 1: $c_{1,5}$ is a Gumbel copula with $\tau_{1,5}=0.6$, $c_{2,4}$ is a BB1 copula with $\tau_{2,4}=0.83$, $c_{3,4}$ is a BB7 copula with $\tau_{3,4}=0.74$ and $c_{4,5}$ is a Tawn copula with $\tau_{4,5}=0.72$; 
	 \item Tree 2: $c_{1,4;5}$ is a Clayton copula with $\tau_{1,4;5}=0.5$, $c_{2,5;4}$ is a Joe copula with $\tau_{2,5;4}=0.45$ and $c_{3,5;4}$ is a BB6 copula with $\tau_{3,5;4}=0.48$; 
	 \item Tree 3: $c_{1,3;4,5}$ is a t copula with $\tau_{1,3;4,5}=-0.19$ and $\nu_{1,3;4,5}=3$ degrees of freedom and $c_{2,3;4,5}$ is a Frank copula with $\tau_{2,3;4,5}=-0.31$; 
	 \item Tree 4: $c_{1,2;3,4,5}$ is a Gaussian copula with $\tau_{1,2;3,4,5}=-0.13$.
\end{itemize}

As described above, we perform the following steps 100 times: Generate a sample of size $N=3000$ from the specified vine copula and fit four different models to the data sample (a Gaussian copula, a C-vine, a D-vine and an R-vine). \autoref{tab:ModSelMixed5d} displays the number of parameters, the dKL to the five-dimensional independence copula and the AIC and BIC values of the four fitted models, all averaged over the 100 replications. The corresponding estimated standard errors are given in brackets.
\begin{table}[h!]
	\centering
	\begin{tabular}{lrrrr}
		\hline
		& \# par& $\dKL(\,\cdot\,,c^\perp)$ & AIC & BIC  \\ 
		\hline
		Gaussian copula & $\mathbf{10.00}\, (0.00)$ & $6.24\, (0.09)$ & $-23669\, (377)$ & $-23609\, (377)$ \\ 
		C-vine & $15.57\, (0.64)$ & $7.10\, (0.06)$ & $-28317\, (333)$ & $-28223\, (333)$ \\ 
		D-vine & $19.77\, (0.47)$ & $7.41\, (0.08)$ & $-30320\, (354)$ & $-30201\, (354)$ \\ 
		R-vine & $15.68\, (0.78)$ & $\mathbf{8.37}\, (0.06)$ & $\mathbf{-33843}\, (344)$ & $\mathbf{-33749}\, (345)$ \\ 
		\hline
	\end{tabular}
	\caption{Average number of parameters, dKL to the five-dimensional independence copula, AIC and BIC of the fitted Gaussian copula, C-vine, D-vine and R-vine. Standard errors are given in brackets. Best values per column are marked in bold.}
	\label{tab:ModSelMixed5d}
\end{table}

  Compared to the 15 parameters of the true model, the Gaussian copula has only 10 parameters but also exhibits the poorest fit of all considered models with respect to any of the decision criteria. The C-vine (between 15 and 16 parameters on average) is ranked third by dKL, AIC and BIC. The D-vine model uses the most parameters (almost 20) but also performs better than the C-vine. With just under 16 parameters on average the R-vine copula is rated best by all three measures. We see that the ranking of the four fitted models is the same for dKL, AIC and BIC. We also checked that all 100 cases yielded this ranking. Considering the empirical `noise-to-signal' ratio, i.e. the quotient of the standard errors and the absolute estimated mean, we obtain that the dKL performs better than AIC and BIC (e.g.\ for the R-vine we have $0.06/8.37<344/33843<345/33479$).

\subsection{20-dimensional t vine}\label{sec:20dtvine}
In order to show a high-dimensional example, we consider a 20-dimensional D-vine being also a t vine, i.e.\ a vine copula with only bivariate t copulas as pair copulas. The association parameter is chosen constant for all pair-copulas in one tree: Kendall's $\tau$ in Tree $m$ is $0.8^m$, $m=1,\ldots,19$. Further, all pairs are heavy-tailed, having $\nu=3$ degrees of freedom. Due to the overall constant degrees of freedom the resulting t vine with its 380 parameters is not a t copula (cf.\ \autoref{sec:vinecopulas}). Now we repeat the following procedure 100 times: Generate a sample of size $N=3000$ from the t vine and fit a Gaussian copula, a t copula, a t vine and an R-vine with arbitrary pair-copula families to the simulated data. Since the calculation of the dKL in $d=20$ dimensions would be rather time-consuming, we use the sdKL instead. We present the number of parameters, the sdKL to the 20-dimensional independence copula and the AIC and BIC values of the four fitted models (again averaged over the 100 replications) in \autoref{tab:ModSelT20d}. The estimated standard errors are given in brackets. 
\begin{table}[h!]
	\centering
	\begin{tabular}{lrrrr}
		\hline
		& \# par & $\sdKL(\,\cdot\,,c^\perp)$ & AIC & BIC \\ 
		\hline
		Gaussian copula & $\mathbf{190.00}\, (0.00)$ & $84.11 \, (0.75) $ & $-271610\, (2962)$ & $-270468\, (2962)$ \\ 
		t copula & $191.00\, (0.00)$ & $93.95\, (0.67)$ & $-299703\, (1929)$ & $-298556\, (1929)$ \\ 
		t vine & $380.00\, (0.00)$ & $96.72\, (0.82)$ & $\mathbf{-309647}\, (2112)$ & $\mathbf{-307365}\, (2112)$ \\ 
		R-vine & $379.87\, (0.60)$ & $\mathbf{96.80}\, (0.99)$ & $-309337\, (2579)$ & $-307056\, (2579)$ \\ 
		\hline
	\end{tabular}
	\caption{Average number of parameters, sdKL to the 20-dimensional independence copula, AIC and BIC of the fitted Gaussian copula, t copula, t vine and R-vine. Standard errors are given in brackets. Best values per column are marked in bold.}
	\label{tab:ModSelT20d}
\end{table}

The Gaussian copula has the least parameters (190) but also the worst sdKL, AIC and BIC values. Adding a single additional parameter already causes an enormous improvement of all three measures for the t copula. The t vine is more flexible but has considerably more parameters than the t copula (380); nevertheless all three decision criteria prefer the t vine over the t copula. Surprisingly, the t vine is even ranked a little bit higher by AIC and BIC than the R-vine, which also has roughly 380 parameters on average. This ranking might seem illogical at first because the class of R-vines is a superset of the class of t vines such that one would expect the fit of the R-vine to be at least as good as the fit of the t vine. The reason for this alleged contradiction is that the fitting procedure that is implemented in the R package \texttt{VineCopula} \citep{VC} is not optimizing globally but tree-by-tree (cf.\ \autoref{sec:vinecopulas}). Therefore, it is possible that fitting a non-t copula in one of the lower trees might be optimal but cause poorer fits in some of the higher trees. However, the difference between the fit of the t vine and the R-vine is very small and for 83 of the 100 samples both procedures fit the same model. Therefore, we want to test whether this difference is significant at all. For this purpose, we perform a parametric bootstrapping based test as described in \autoref{sec:paramboots} at the level $\alpha=5\%$ with $M=100$ replications. With p-values between $0.32$ and $0.78$ we cannot even reject the null hypothesis that the two underlying models coincide in any of the remaining 17 cases, where different models were fitted. Hence we would prefer to use the simpler t vine model which is in the same model class as the true underlying model. In a similar manner we check whether the difference between the t copula and the t vine is significant. Here, however, we find out that the model can indeed be distinguished at a $5\%$ confidence level for all 100 samples (p-values range between $0$ and $0.03$). Considering the empirical noise-to-signal ratio we see that sdKL is a bit more dependent on the sample compared to dKL such that AIC, BIC and sdKL have roughly the same noise-to-signal ratio, where the values of AIC/BIC are slightly lower for the t copula, the t vine and the R-vine.

\section{Determination of the optimal truncation level}\label{sec:truncation}

As the dimension $d$ of a vine copula increases, the number of parameters $d(d-1)/2$ grows quadratically. For example, a 50-dimensional R-vine consists of 1225 (conditional) pair-copulas, each with one or more copula parameters. This on the one hand can create computational problems, while on the other hand the resulting model is difficult to interpret. Given an $d$-dimensional data set ($d$ large), it has been proposed \cite[see][]{brechmann2012truncated,brechmann2015truncation} to fit a so-called $k$-truncated vine, where the pair-copulas of all trees of higher order than some truncation level $k\leq d-1$ are modeled as independence copulas. This reduces the number of pair-copulas to be estimated from $d(d-1)/2$ to $k(k-1)/2$, where $k$ is chosen as small as can be justified. The heuristic behind this approach is that the sequential fitting procedure of regular vines captures most of the dependence in the lower trees, such that the dependence in the higher trees might be negligible and therefore the approximation error caused by using an independence copula is rather small. The task of finding the optimal truncation level $k^*$ has already been tackled in the recent literature. \cite{brechmann2012truncated} use likelihood based criteria such as the AIC, BIC and Vuong test for the selection of $k^*$, while \cite{brechmann2015truncation} propose an approach based on fit indices that measure the distance between fitted and observed correlation matrices.
\subsection{Algorithms for the determination of optimal truncation levels}
Using the proposed distance measures we can directly compare several truncated vines with different truncation levels. With the bootstrapped confidence intervals described in \autoref{sec:paramboots} we can assess whether the distances are significant in order to find the optimal truncation level. To be precise, in the following we present two algorithms that use the sdKL for the determination of the optimal truncation level, a global one (\autoref{Alg1}) and a sequential one (\autoref{Alg2}).

In \autoref{Alg1}, tRV($k$) denotes the $k$-truncated version of RV. Since a full $d$-dimensional R-vine consists of $d-1$ trees, tRV($d-1$) and RV coincide.

\begin{algorithm}
	\caption{Global determination of the optimal truncation level}
	\label{Alg1}
	\textbf{Input:} $d$-dimensional copula data, significance level $\alpha$.
	\begin{algorithmic}[1]
		\State Fit full (non-truncated) regular vine $\text{RV}=\text{tRV}($d-1$)$ to the data set.
		\For {$m=d-2,\ldots,0$}
		\State \parbox[t]{\dimexpr\linewidth-\algorithmicindent}{Specify the truncated vine tRV($m$) by setting all pair-copulas of trees $m+1$ and higher to the independence copula.\strut}
		\State \parbox[t]{\dimexpr\linewidth-\algorithmicindent}{Calculate the sdKL between RV and tRV($m$) and use the parametric bootstrap to check whether the distance is significantly different from zero.\strut}
		\If {distance is significant}
		\State \textbf{break} the \textbf{for}-loop and \textbf{return} the optimal truncation level $k^*=m+1$.
		\EndIf
		\EndFor
	\end{algorithmic}
	\textbf{Output:} Optimal truncation level $k^*=m+1$.
\end{algorithm}
The algorithm starts with the full model RV and, going backwards, truncates the vine tree-by-tree until the distance between the $m$-truncated vine and the full model is significantly larger than 0. Hence, the truncation at level $m$ is too restrictive such that we select the level $k^*=m+1$, for which the distance was still insignificant. For the testing procedure we can use the test from \autoref{sec:paramboots} since the class of truncated vine copula models is nested in the general class of all vine copulas.

Since fitting a full vine copula model might be rather time-consuming in high dimensions 
%Since one of the main advantages of considering truncated vines is reduced numerical complexity of fitting, one might argue that this way of finding the optimal truncation level for a high-dimensional data set defeats its purpose because in the process the full R-vine has to be estimated. Therefore, 
with \autoref{Alg2} we propose another procedure of determining the truncation level, which builds the R-vine sequentially tree by tree, starting with the first tree. In each step we check whether the additionally modeled tree significantly changes the resulting model in comparison to the previous one. As long as it does, the vine is updated to one with an additionally modeled tree. Only when the addition of a new tree of order $m$ results in a model that is statistically indistinguishable from the previous one, the algorithm stops and returns the optimal truncation level $k^*=m-1$.

\begin{algorithm}
	\caption{Sequential determination of the optimal truncation level}
	\label{Alg2}
	\textbf{Input:} $d$-dimensional copula data, significance level $\alpha$.
	\begin{algorithmic}[1]
		\State Set tRV(0) to be a truncated vine with truncation level 0, i.e.\ an independence copula.
		\For {$m=1,\ldots,d-1$}
		\State \parbox[t]{\dimexpr\linewidth-\algorithmicindent}{Specify the truncated vine tRV($m$) by taking the truncated vine from the previous step tRV($m-1$) and estimating the pair-copulas from tree $m$.\strut}
		\State \parbox[t]{\dimexpr\linewidth-\algorithmicindent}{Calculate the sdKL between tRV($m-1$) and tRV($m$) and use the parametric bootstrap to check whether the distance is significantly different from zero.\strut}
		\If {distance is not significant}
		\State \textbf{break} the \textbf{for}-loop and \textbf{return} the optimal truncation level $k^*=m-1$
		\EndIf
		\EndFor
	\end{algorithmic}
	\textbf{Output:} Optimal truncation level $k^*=m-1$
\end{algorithm}
The heuristic behind \autoref{Alg2} is that since the vine is estimated sequentially maximizing the sum of absolute (conditional) Kendall's $\tau$ values in each tree \citep[for details see][]{dissmann2013selecting}, we can expect the distance between two subsequent truncated vines to be decreasing. Therefore, if the distance between tRV($k^*$) and tRV($k^*+1$) is not significant, the distances between tRV($m$) and tRV($m+1$) for $m>k^*$ should be not significant either.

Comparing \autoref{Alg1} and \autoref{Alg2}, we note that in general they do not find the same truncation level. For example, consider the case where for some $m$ the distances between tRV($m$) and tRV($m+1$), tRV($m+1$) and tRV($m+2$), until tRV($d-2$) and tRV($d-1$) are not significant, while the distance between tRV($m$) and tRV($d-1$) is. Then, \autoref{Alg2} would return an optimal truncation level of $m$, whereas we would obtain a higher truncation level by \autoref{Alg1}. So in general we see that \autoref{Alg2} finds more parsimonious models than \autoref{Alg1}.

In the following we examine how well the proposed algorithms for finding optimal truncation levels for R-vines work in several simulated scenarios as well as real data examples. We compare our results to the existing methodology of \cite{brechmann2012truncated}, who use a Vuong test (with and without AIC/BIC correction) to check whether there is a significant difference between a certain $k$-truncated vine and the corresponding vine with truncation level $k-1$, for $k=1,\ldots,d-1$. Starting with the lowest truncation levels, once the difference is not significant for some $m$, the algorithm stops and returns the optimal truncation level $k^*=m-1$.

\subsection{Simulation study}

\begin{figure}[htbp]
	\centering
	\includegraphics[width=0.4\textwidth]{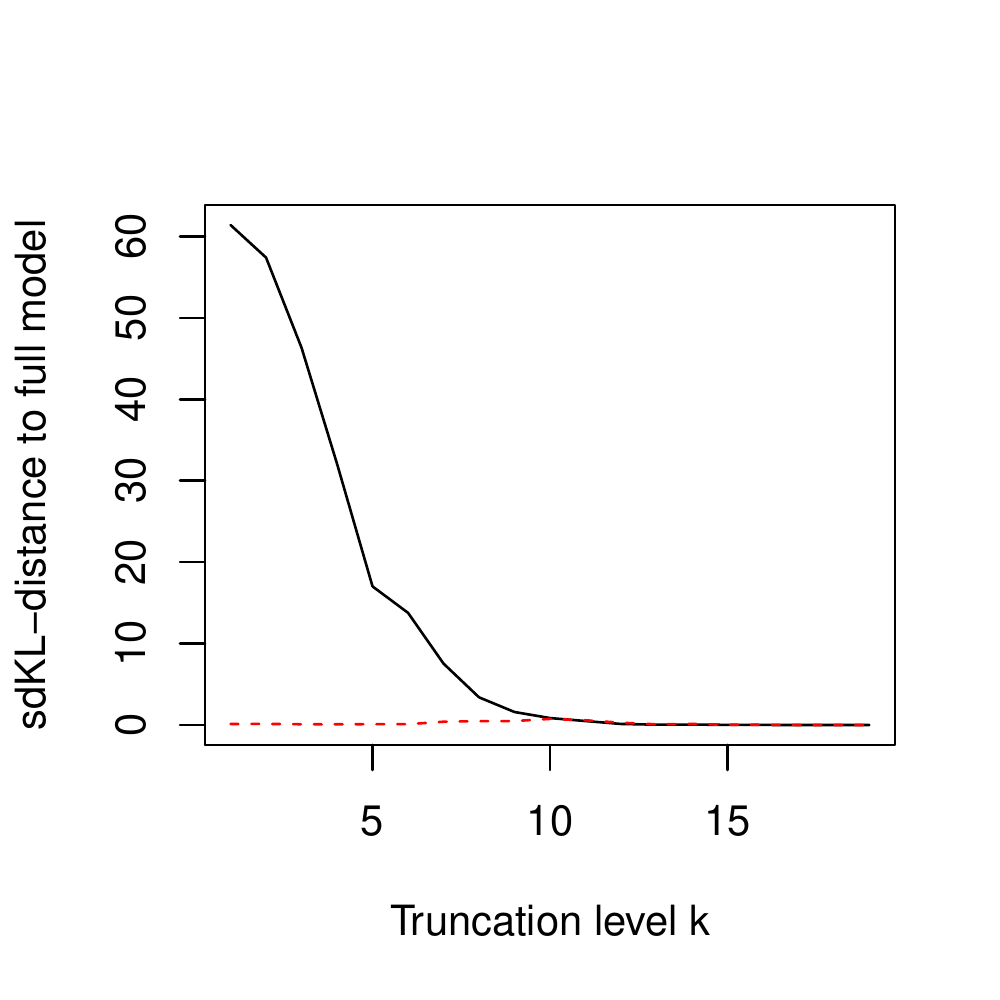}
	\includegraphics[width=0.4\textwidth]{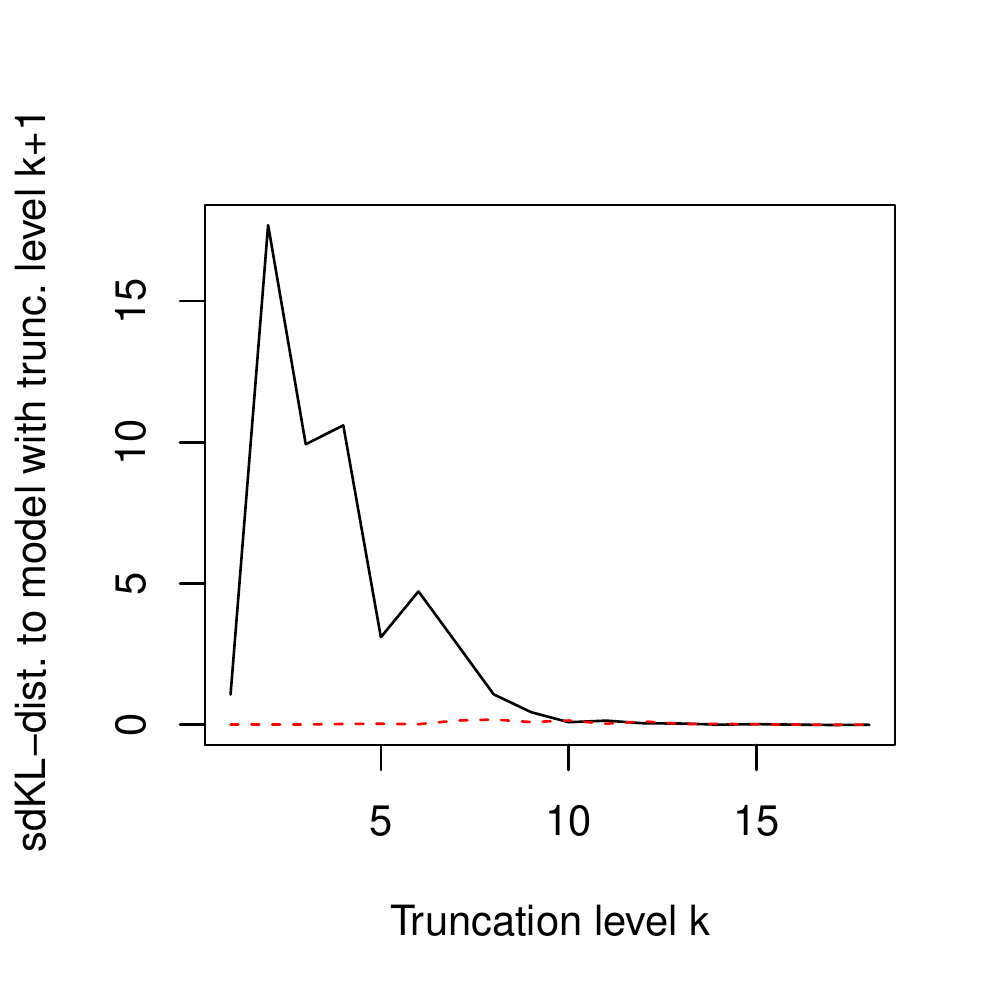}
	\caption{Visualization of the algorithms for data generated from a 20-dimensional t copula truncated at level 10.\\Left (\autoref{Alg1}):\ sdKL-distance to full model with dashed bootstrapped 95\% confidence bounds.\\Right (\autoref{Alg2}):\ sdKL-distance to model with truncation level $k+1$ with dashed bootstrapped 95\% confidence bounds.}
	\label{fig:smooth10trunc}
\end{figure}

\paragraph{20-dimensional t copula truncated at level 10}
In the first simulated example, we consider a scenario where the data comes from a 20-dimensional t copula truncated at level 10. For this, we set the degrees of freedom to 3 and produce a random correlation matrix sampled from the uniform distribution on the space of correlation matrices \citep[as described in][]{joe2006generating}. In this example, the resulting correlations range between $-1$ and $1$ with a higher concentration on correlations with small absolute values. After sampling the correlation matrix, we express the corresponding t copula as a D-vine (cf.\ \autoref{sec:vinecopulas}) and truncate it at level 10, i.e.\ the pair-copulas of trees 11 to 19 are set to the independence copula. From this truncated D-vine we generate a sample of size $N=2000$ and use the R function \texttt{RVineStructureSelect} from the package \texttt{VineCopula} to fit a vine copula to the sample with the Di\ss mann algorithm (see \autoref{sec:vinecopulas}). The question is now if our algorithms can detect the true truncation level underlying the generated data. For this we visualize the steps of the two algorithms. Concerning \autoref{Alg1}, in the left panel of \autoref{fig:smooth10trunc} we plot the sdKL-distances between the truncated vines and the full (non-truncated) vine against the 19 truncation levels together with the bootstrapped 95\% confidence bounds ($d_{95}$ from \autoref{sec:paramboots}) under the null hypothesis that the truncated vine coincides with the full model (dashed line).

Naturally, the curve corresponding to \autoref{Alg1} is decreasing with an extremely large distance between the one-truncated vine and the full model and a vanishingly small distance between the 18-truncated vine and the full model, which only differ in the specification of one pair-copula. In order to determine the smallest truncation level whose distance to the full model is insignificantly large, the algorithm compares these distances to the bootstrapped 95\% confidence bounds. In this example we see that the smallest truncation level for which the sdKL-distance to the full model drops below the confidence bound is 10, such that the algorithm is able to detect the true truncation level. In order to check, whether this was not just a coincidence we repeated this procedure 100 times and found that the optimal truncation level found by the algorithm averages to 10.5 with a standard deviation of 0.81.

The right panel of \autoref{fig:smooth10trunc} displays the results for \autoref{Alg2}. For each truncation level $k$, the sdKL-distance between the vine truncated at level $k$ and the vine truncated at level $k+1$ is plotted, again together with bootstrapped 95\% confidence bounds under the null hypothesis that this distance is 0, i.e.\ the true model is the one with truncation level $k$. We observe that the largest sdKL-distance is given between the vine copulas truncated at levels 2 and 3, 3 and 4, and 4 and 5, respectively. This is in line with the results from \autoref{Alg1} (left panel of \autoref{fig:smooth10trunc}), where we observe the steepest decrease in sdKL to the full model from truncation level 2 to 5. In this example \autoref{Alg2} would also detect the true truncation level 10. In the 100 simulated repetitions of this scenario, the average optimal truncation level was 10.2 with a standard deviation of 0.41. 

In each of the 100 repetitions, we also used the Vuong test based algorithms without/with AIC/with BIC correction from \cite{brechmann2012truncated} to compare our results. They yielded average truncation levels of 14.6, 12.6 and 10.8 (without/with AIC/with BIC correction), depending on the correction method. So all three methods overestimate the truncation level, in particular the first two.

Thus we have seen that in a scenario where the data is generated from a truncated vine both our proposed algorithms manage to detect the truncation level very well. Next, we are interested in the results of the algorithms when the true underlying copula is not truncated.

\begin{figure}[!htb]
	\centering
	\includegraphics[width=0.4\textwidth]{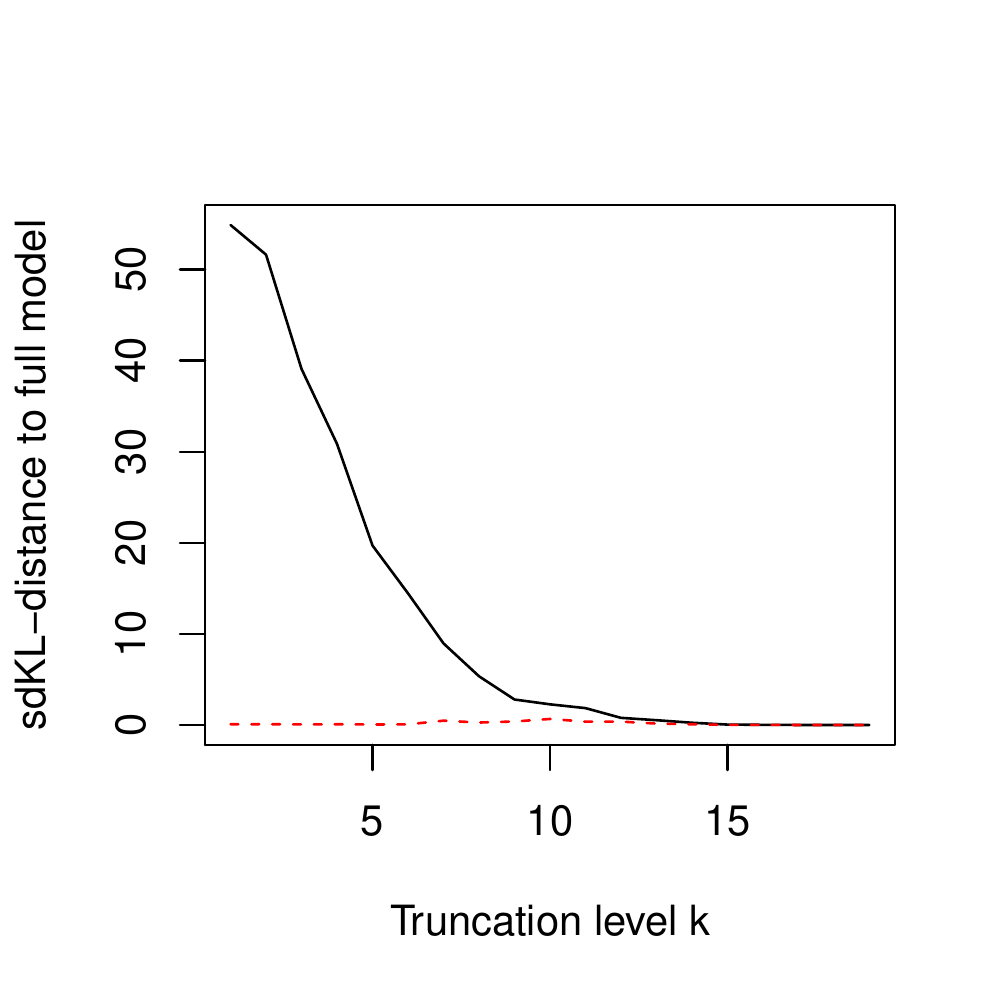}
	\includegraphics[width=0.4\textwidth]{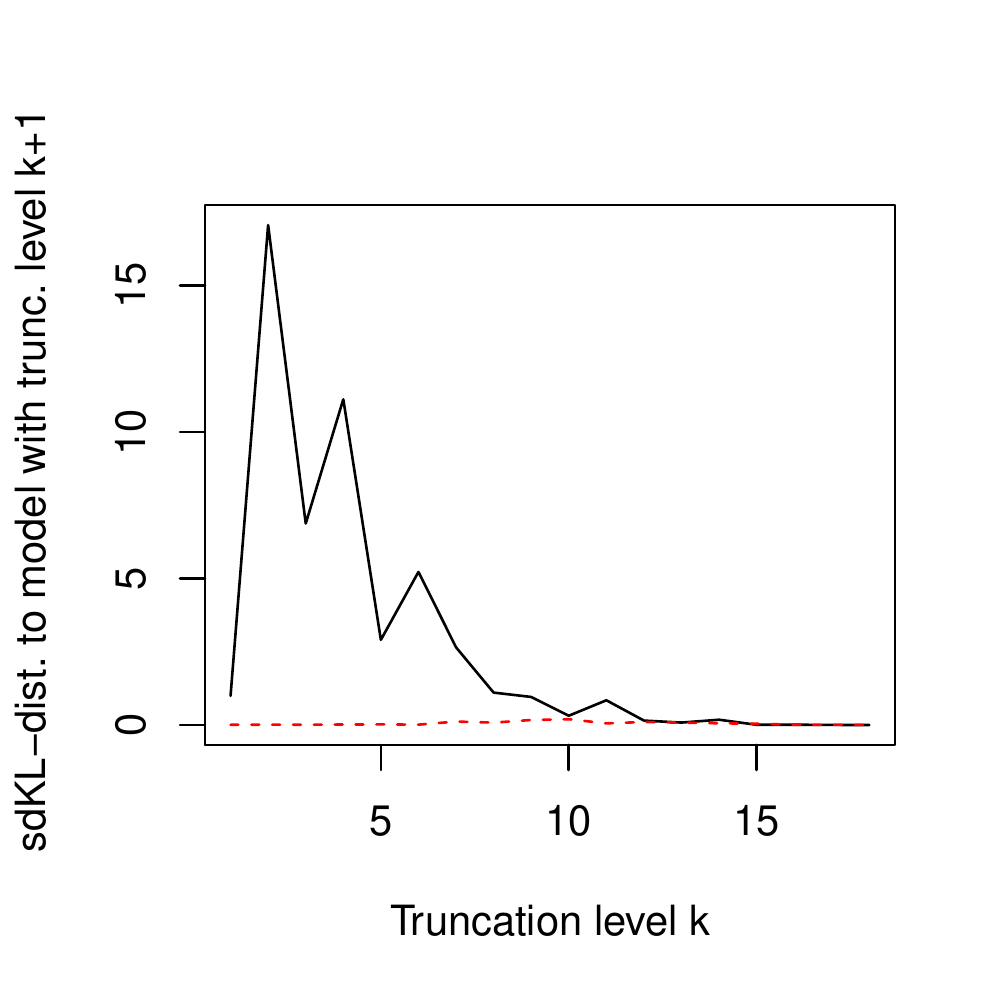}
	\caption{Visualization of the algorithms for data generated from a 20-dimensional t copula (non-truncated).\\Left (\autoref{Alg1}):\ sdKL-distance to full model with dashed bootstrapped 95\% confidence bounds.\\Right (\autoref{Alg2}):\ sdKL-distance to model with truncation level $k+1$ with dashed bootstrapped 95\% confidence bounds.}
	\label{fig:smooth0trunc}
\end{figure}

\paragraph{20-dimensional t copula (non-truncated)}
In this example we generate data from the same 20-dimensional t copula as before, this time without truncating it. The results of the algorithms are displayed in \autoref{fig:smooth0trunc}.

At first sight the plots look quite similar to those of \autoref{fig:smooth10trunc}. Due to the sequential fitting algorithm of \cite{dissmann2013selecting}, which tries to capture large dependencies as early as possible (i.e.\ in the lower trees), the sdKL distance to the full model (left panel of \autoref{fig:smooth0trunc}) is strongly decreasing in the truncation level. However, for truncation levels 10 to 15 this distance is still significantly different from zero (albeit very close to the 95\% confidence bounds for $k\geq 12$) such that the optimal truncation level is found to be 16. The right panel of \autoref{fig:smooth0trunc} tells us that the distance between the 11- and 12-truncated vine copulas is still fairly large and all subsequent distances between the $k$- and $(k+1)$-truncated models are very small. However, \autoref{Alg2} also detects 16 to be the optimal truncation level because the distances are still slightly larger than the 95\% confidence bounds for smaller $k$. In the 100 simulated repetitions the detected optimal truncation level was between 14 and 18 with an average of 16.2 for \autoref{Alg1} and 15.4 for \autoref{Alg2}. 

Again, we used the algorithms from \cite{brechmann2012truncated} in each of the 100 repetitions. From the different correction methods we obtained the following average truncation levels: 18.6, 18.3 and 17.6 (without/with AIC/with BIC correction). 

Hence we can conclude that for vine copulas fitted by the algorithm of \cite{dissmann2013selecting} our algorithms decide for a little more parsimonious models than the ones from \cite{brechmann2012truncated}. This can even be desirable since the fitting algorithm by \cite{dissmann2013selecting} selects vine copulas such that there is only little strength of dependence in high-order trees. Therefore, we do not necessarily need to model all pair-copulas of the vine specification explicitly and a truncated vine often suffices.
%\paragraph{20-dim t copula with rapidly decreasing dependence}
%In the third example the data comes from a non-truncated 20-dimensional t-copula whose (conditional) dependencies are rapidly decreasing from tree to tree.
%\begin{figure}[!htb]
%\centering
%\includegraphics[width=0.4\textwidth]{Truncation_Plots09.pdf}
%\includegraphics[width=0.4\textwidth]{Truncation_Plots10.pdf}
%\caption{Simulation from a 20-dim t copula with rapidly decreasing dependence (non-truncated).\\Left: sdKL-distance to full model.\\Right: sdKL-distance to model with truncation level k+1}
%\label{fig:trunc4}
%\end{figure}

\subsection{Real data examples}\label{sec:trunc_real_data}

Having seen that the algorithms seem to work properly for simulated data we now want to turn our attention to real data examples. First we revisit the example considered in \cite{brechmann2012truncated} concerning 19-dimensional Norwegian finance data.

\paragraph{19-dimensional Norwegian finance data}

\begin{figure}[!htb]
	\centering
	\includegraphics[width=0.4\textwidth]{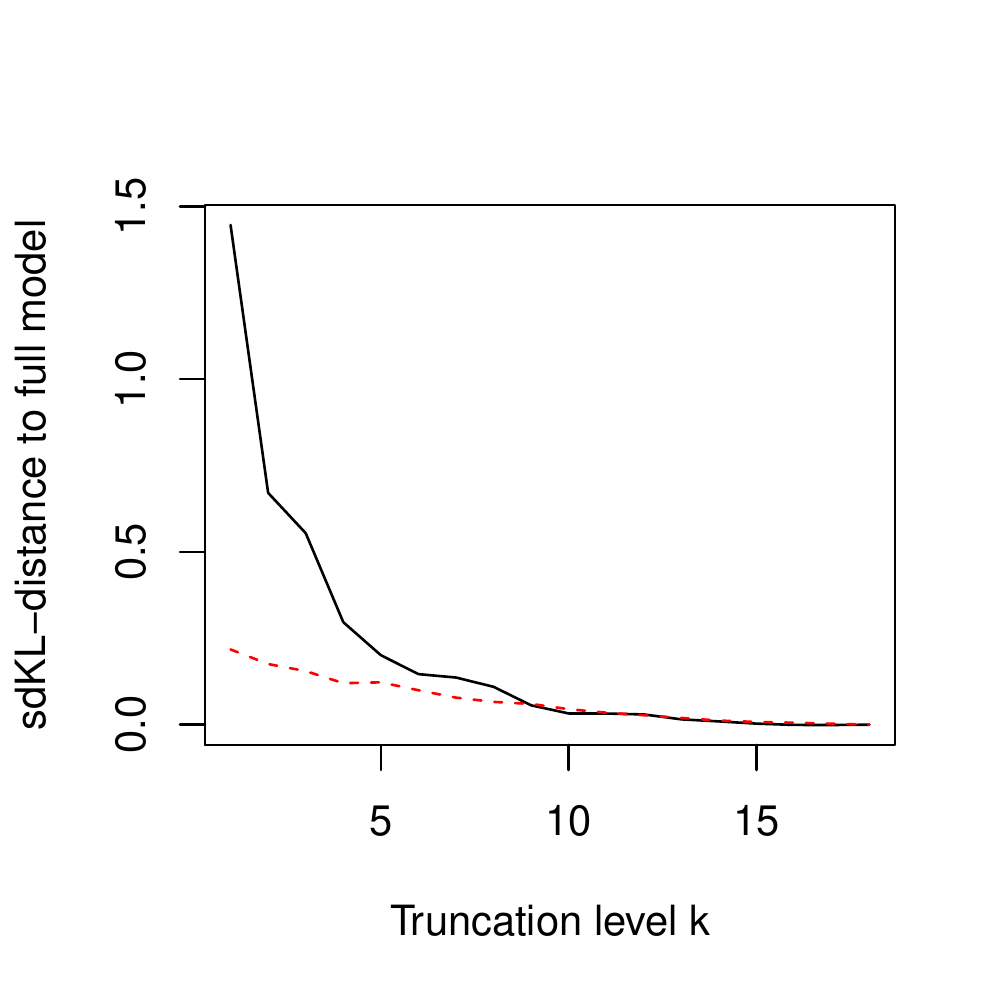}
	\includegraphics[width=0.4\textwidth]{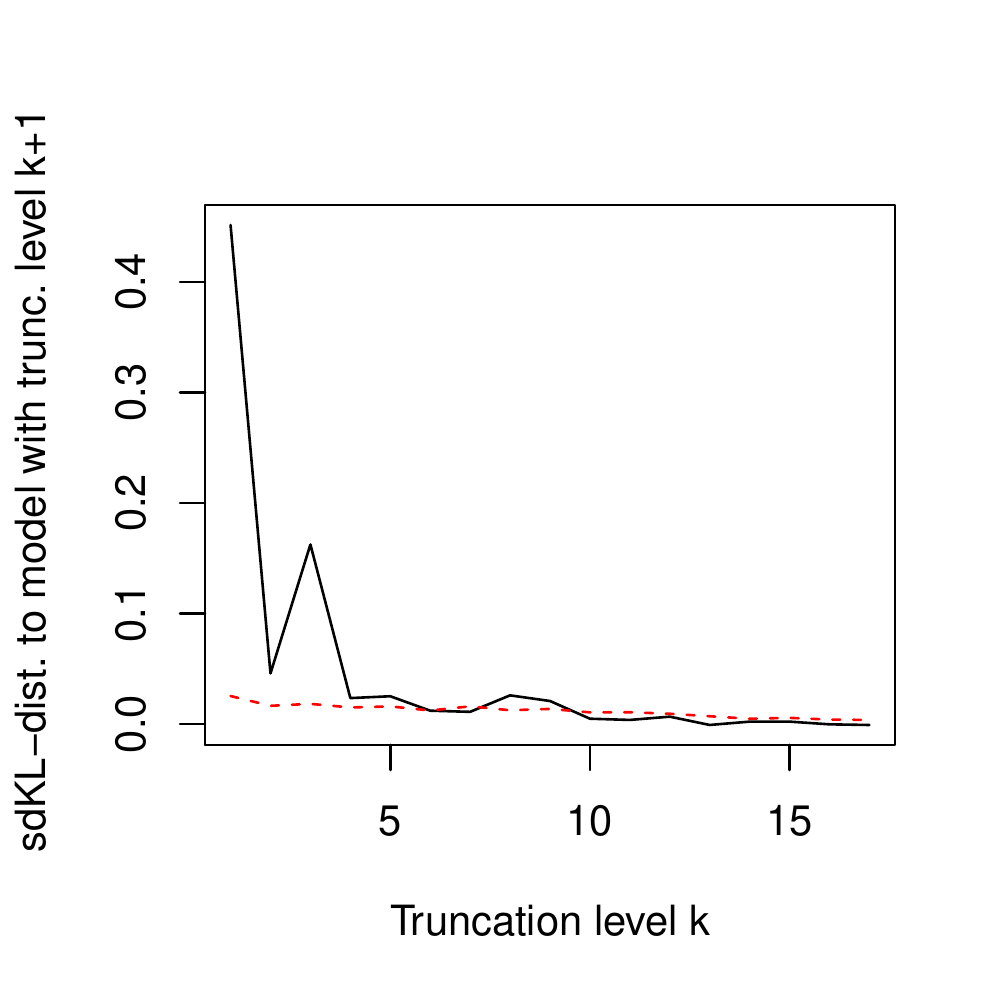}
	\caption{Visualization of the algorithms for the 19-dimensional Norwegian finance data.\\Left (\autoref{Alg1}):\ sdKL-distance to full model with dashed bootstrapped 95\% confidence bounds.\\Right (\autoref{Alg2}):\ sdKL-distance to model with truncation level $k+1$ with dashed bootstrapped 95\% confidence bounds.}
	\label{fig:oslo}
\end{figure}

The data set consists of 1107 observations of 19 financial quantities such as interest rates, exchange rates and financial indices for the period 2003--2008 \cite[for more details refer to][]{brechmann2012truncated}. \autoref{fig:oslo} shows the visualization of the two algorithms for this data set.

We see that the sdKL-distance to the full model is rapidly decreasing in the truncation level $k$, being quite close to the 95\% confidence bound for $k\geq4$, very close for $k\geq6$ and dropping below it for $k=10$. Hence we can conclude that the optimal truncation level found by \autoref{Alg1} is 10, while a truncation level of 6 or even 4 may also be justified if one seeks more parsimonious models. This is exactly in line with the findings of \cite{brechmann2012truncated}, who ascertained that depending on the favored degree of parsimony both truncation levels 4 and 6 may be justified. Yet, they find that there still are significant dependencies beyond the sixth tree. This can also be seen from the right plot of \autoref{fig:oslo}, which visualizes the results from \autoref{Alg2}. We see that the distance between two subsequent truncated vines first falls below the 95\% confidence bound for $k=6$, after being close to it for $k=4$ and $k=5$. Thus we see that in this example \autoref{Alg2} indeed finds a more parsimonious model than \autoref{Alg1}. If we took the distances between all subsequent truncated vines into account, we would see that trees 9 and 10 still contribute significant dependencies, such that the ``global'' optimal truncation level again would be 10. If a data analyst decided that the parsimonious model truncated at level 6 or 4 would suffice for modeling this 19-dimensional data set, he or she would be able to reduce the number of pair-copulas to be modeled from 171 of the full model to 93 or 66, respectively, and thus greatly improve model interpretation and simplify further computations involving the model (e.g.\ Value-at-Risk simulations).

\begin{figure}[!htb]
	\centering
	\includegraphics[width=0.4\textwidth]{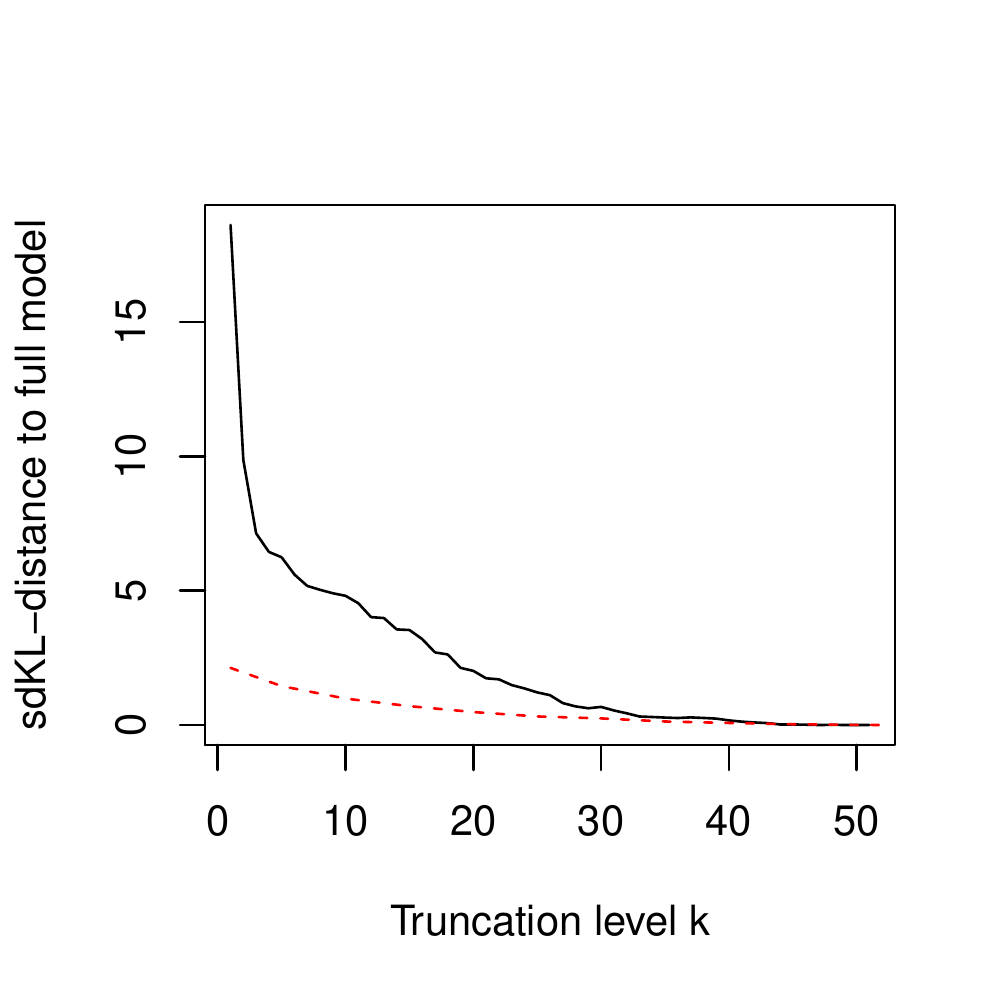}
	\includegraphics[width=0.4\textwidth]{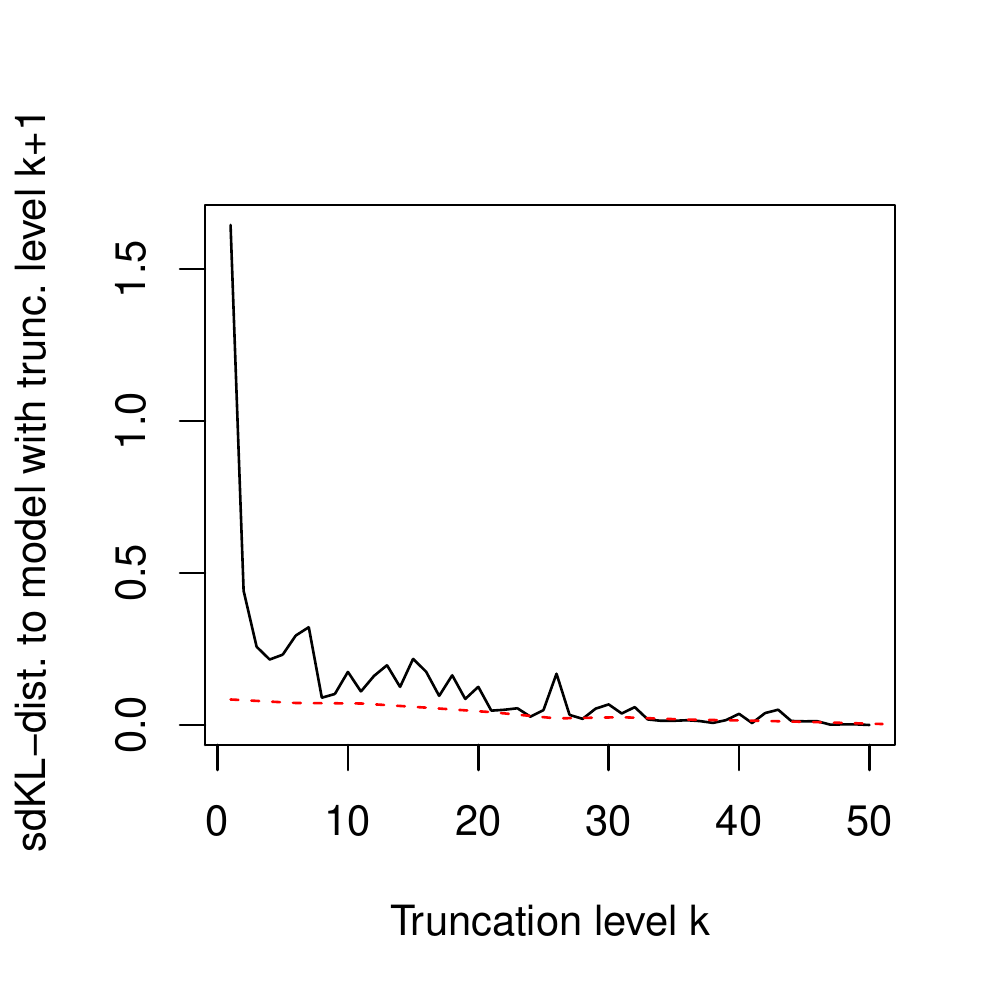}
	\caption{Visualization of the algorithms for the 52-dimensional EuroStoxx50 data.\\Left (\autoref{Alg1}):\ sdKL-distance to full model with dashed bootstrapped 95\% confidence bounds.\\Right (\autoref{Alg2}):\ sdKL-distance to model with truncation level $k+1$ with dashed bootstrapped 95\% confidence bounds.}
	\label{fig:ES50}
\end{figure}

\paragraph{52-dimensional EuroStoxx50 data}
Since the positive effect of truncating vine copulas intensifies with increasing dimensions, we revisit the 52-dimensional EuroStoxx50 data set from \autoref{sec:satestExamples}. For risk managers it is an relevant task to correctly assess the interdependencies between these variables since they are included in most international banking portfolios. \autoref{fig:ES50} shows the results of the algorithms for this data set.

In the left panel we see that most of the dependence is captured by the first few trees since there the sdKL-distance to the full model has its sharpest decrease in truncation level $k$. The distance gets very close to the dashed 95\% confidence bound for $k>27$, however crossing it not before $k=43$, implying a rather high truncation level. Considering the visualized results of \autoref{Alg2} in the right panel of \autoref{fig:ES50} we observe that the distances between subsequent truncated vines is quite small for $k\geq 8$, first dropping below the 95\% confidence bound for $k=24$. However, it increases again afterwards and ultimately drops below the confidence bound for $k=44$. One could argue that a truncation level of $k=33$ might be advisable since the treewise distance slightly exceeds the confidence bound only three times thereafter. This would reduce the number of pair-copulas to be modeled from 1326 for the full 52-dimensional model to 1155 for the 33-truncated vine copula. Thus, with the help of model distances we can find simpler models for high-dimensional data. 

For comparison, the algorithm from \cite{brechmann2012truncated} finds optimal truncation levels of 47 (without correction), 24 (AIC correction) and 3 (BIC correction). We see that there are large differences between the three methods: Whereas a truncation level of 47 corresponds almost to the non-truncated vine, one should be skeptical whether a 3-truncated vine is apt to describe the dependence structure of 52 random variables.

\section{Conclusion}\label{sec:conclusion}

Vine copulas are a state-of-the-art method to model high-dimensional copulas. The applications presented in this paper show the necessity of calculating distances between such high-dimensional vine copulas. In essence, whenever we have more than one vine model to describe observed data, be it a simplified and a non-simplified vine, vines with different truncation levels or with certain restrictions on pair-copula families or the underlying vine structure, model distances help to select the best out of the candidate models. The modifications of the Kullback-Leibler distance introduced in \cite{killiches2015model} have proven to be fast and accurate even in high dimensions, where the numerical calculation of the KL is infeasible. While in this paper we only considered datasets with dimensions $d\leq52$, applications in even higher dimensions are possible. With the theory developed in \cite{muller2016representing} the fitting of vines with hundreds of dimensions is facilitated with the focus on sparsity, i.e.\ fitting as many independence copulas as justifiable in order to reduce the number of parameters. In ongoing research the proposed distance measures are applied to select between several of these high-dimensional models.

\section*{Acknowledgment}\label{sec:acknowledgment}
The third author is supported by the German Research Foundation (DFG grant CZ 86/4-1). Numerical calculations were performed on a Linux cluster supported by DFG grant INST 95/919-1 FUGG. 

\bibliographystyle{apalike}
\bibliography{references2}

\end{document}